\documentclass[apj]{emulateapj}
\usepackage{graphicx}
\usepackage{amsmath}
\usepackage[varg]{txfonts}

\newcommand{\hfs}   {\textsl{HfS}}
\newcommand{\hfsnh} {\textsl{HfS\_nh3}}
\newcommand{\nh}    {NH$_3$}
\newcommand{\Nnht}  {N(\mathrm{NH_3})}
\newcommand{\vlsr}  {V_\mathrm{LSR}}
\newcommand{\Tex}   {T_\mathrm{ex}}
\newcommand{\Trot}  {T_\mathrm{rot}}
\newcommand{\Tk}    {T_\mathrm{k}}
\newcommand{\Tbg}   {T_\mathrm{bg}}
\newcommand{\Tnu}   {T_\nu}
\newcommand{\kms}   {km~s$^{-1}$}

\begin{document}

\slugcomment{accepted by PASP}
\date\today

\title{\textsl{H\lowercase{f}S}, Hyperfine Structure Fitting Tool}
\shorttitle{HfS}

\author{Robert Estalella\altaffilmark{\dag}
\email{robert.estalella@ub.edu}}
\shortauthors{R. Estalella}

\affil{
Departament de F\'{\i}sica Qu\`antica i Astrof\'{\i}sica
(formerly Astronomia i Meteorologia), 
Institut de Ci\`encies del Cosmos (ICC),\\ 
Universitat de Barcelona (IEEC-UB), 
Mart\'{\i} i Franqu\`es 1, 
E08028 Barcelona, Spain}

\altaffiltext{\dag}{
The ICC (UB) is a CSIC-Associated Unit through the ICE (CSIC)}

\begin{abstract}
\textsl{HfS} is a tool to fit the hyperfine structure of spectral lines, with
multiple velocity components. 
The \textsl{HfS\_nh3} procedures included in \textsl{HfS} fit simultaneously the
hyperfine  structure of the NH$_3$ $(J,K)=(1,1)$ and $(2,2)$ transitions, and
perform a standard analysis to derive $T_\mathrm{ex}$, NH$_3$ column density,
$T_\mathrm{rot}$, and $T_\mathrm{k}$. \textsl{HfS} uses a Monte Carlo approach
for fitting the line parameters. Especial attention is paid to the derivation of
the parameter uncertainties. \textsl{HfS} includes procedures that make use of
parallel computing for fitting spectra from a data cube.
\end{abstract}


\section{Introduction}

\hfs\ (\textbf{H}yper\textbf{f}ine \textbf{S}tructure) \citep{Est16}
is a tool to fit the hyperfine structure of spectral lines, with the possibility
of fitting simultaneously multiple velocity components. 
The \hfsnh\ procedures included in \hfs\ fit simultaneously the hyperfine
quadrupole and magnetic structure of the \nh\ $(J,K)=(1,1)$ and $(2,2)$
inversion transitions.
The assumptions made by \hfs\ are that
the beam filling factor $f$,
the excitation temperature $\Tex$, 
the hyperfine lines linewidth $\Delta V$, and
the central velocity $\vlsr$ 
are the same for all the hyperfine lines.

For \hfsnh\, these assumptions hold for the hyperfine lines of both \nh\
inversion transitions, $(1,1)$ and $(2,2)$. 
In addition, the results of the fit are used by \hfsnh\ to derive
physical parameters including the excitation temperature, \nh\ column density, 
rotational and kinetic temperature, with the assumption that the emitting 
region is homogeneous along the line of sight.

\hfs\ is written in Fortran 77 and 90/95 (mainly for the dynamical storage of
arrays).  
The graphic interface uses the 
PGplot Graphics Subroutine
Library\footnote{\url{http://www.astro.caltech.edu/~tjp/pgplot/}}, 
and some procedures have a multiprocessor version running under 
Open MPI\footnote{\url{http://www.open-mpi.org}}. 
The interactive procedures are menu driven, allowing to select options with the
keyboard, and for some actions, with the mouse cursor and buttons.

One of the advantages of \hfs\ when compared with other packages performing
hyperfine fitting, like for instance, the widely used CLASS of
GILDAS\footnote{\url{https://www.iram.fr/IRAMFR/GILDAS/}} 
is that it is well documented, easy to
install, and with a simple interface. 
\hfs\ can fit, in a single run, multiple velocity components to the spectra of a
FITS data cube, and obtain the maps of the parameters fitted and derived from
the fit, in a short computing time taking advantage of the multiple processors
of current computers.

A preliminary version of \hfs\ was briefly described and used for fitting 
\nh$(1,1)$ and $(2,2)$ spectra in \citet{San13}.

The structure of the paper is as follows:
The fit parameters used by \hfs\ are described in \S \ref{sfitparam},
the fitting strategy in \S \ref{sfitting},
the calculation of the synthetic spectrum in \S \ref{ssynt},
the line parameters derived from the fit parameters in \S \ref{sderived},
the \nh\ physical parameters derived by \hfsnh\ in \S \ref{sphyspar},
the error estimation for the fit parameters in \S \ref{sconf}
and for the derived parameters in \S \ref{snum}.
A comparison with the results obtained with CLASS and the \citet{Ros08} routine
is presented in \S \ref{scomp}.
The different procedures that compose \hfs\ are described in \S \ref{sproc}.
Finally, in several Appendices we describe 
the requisites and installation instructions 
(Appendices \ref{ainstall} to \ref{amp}),
and examples of use of \hfs, and of input and output files 
(Appendices \ref{afitrun} to \ref{aoutput}).

\section{Fit parameters}
\label{sfitparam}

The general \hfs\ procedures fit simultaneously,
for every velocity component of a transition with hyperfine structure, 
four independent parameters, 
\begin{itemize} 

\item $\Delta V$, hyperfine lines linewidth, assumed to be the
same for all the hyperfine lines,

\item $\vlsr$, main line central LSR velocity, 

\item $A^*_m\equiv A(1-\exp\{-\tau_m\})$, main line peak intensity (for 
hyperfine lines wider than the hyperfine separation and the channel width), 
where $A$ is the amplitude (see \S \ref{sderived}), 

\item $\tau^*_m\equiv 1-\exp\{-\tau_m\}$, where $\tau_m$ is the
optical depth of the main line. 

\end{itemize} 

The \hfsnh\ procedures included in \hfs\ fit simultaneously the hyperfine 
structure of a pair of spectra of the \nh\ inversion transitions $(J,K)=(1,1)$
and $(2,2)$ (see Table \ref{thyp}).

Differences in
central velocity of the order of a tenth of km~s$^{-1}$ are usually found
between the $(1,1)$ and $(2,2)$ emissions \citep[see for instance][]{Sep11}. 
Thus, a different $V_\mathrm{LSR}$ is fitted for the $(1,1)$ and the $(2,2)$
spectra, resulting in
two additional parameters for the $(2,2)$ transition, fitted simultaneously
with the four parameters for the $(1,1)$ transition,
$\Delta V$, $V_\mathrm{LSR1}$, $A^*_{1m}$, $\tau^*_{1m}$, 
already described,
\begin{itemize} 

\item $V_\mathrm{LSR2}$, central LSR velocity of the $(2,2)$ transition,

\item $A^*_{2m}\equiv A(1-\exp\{-\tau_{2m}\})$, peak intensity of the
$(2,2)$ main line (for hyperfine lines wider than the hyperfine
separation and the channel width). 

\end{itemize} 
For each set of fit parameters, the optical depth of the $(2,2)$ main line
is obtained from the relation
\begin{equation}
\tau^*_{2m} \equiv 1-\exp\{-\tau_{2m}\} = \tau^*_{1m} \frac{A^*_{2m}}{A^*_{1m}}.
\end{equation}

\begin{table}[htb]
\small
\centering
\caption{\label{thyp}
Relative velocity $V_\mathrm{hyp}$,
and optical depth $\tau_\mathrm{hyp}$ with respect to that of the main line,
$\tau_m$,
of the hyperfine lines of the NH$_3$ $(J,K)=(1,1)$ and $(2,2)$ inversion
transitions \citep{Man15}.
(os: outer satellite, is: inner satellite, m: main)}
\begin{tabular}{lrccrc} 
\hline\hline
&
\multicolumn{2}{c}{$(J,K)=(1,1)$} &&
\multicolumn{2}{c}{$(J,K)=(2,2)$} \\ 
\cline{2-3}\cline{5-6}
&
\multicolumn{1}{c}{$V_\mathrm{hyp}$} & & &
\multicolumn{1}{c}{$V_\mathrm{hyp}$} & \\
&
(km s$^{-1}$) & $\tau_\mathrm{hyp}$ &&
(km s$^{-1}$) & $\tau_\mathrm{hyp}$ \\
\hline
os 
& $-19.548593$ & 0.148148 &  & $-26.557749$ & 0.004186 \\   
& $-19.409429$ & 0.074074 &  & $-26.042011$ & 0.037674 \\   
&              &          &  & $-25.981267$ & 0.020930 \\   
\hline                                                      
is                                                          
& $ -7.815393$ & 0.166667 &  & $-16.401673$ & 0.037209 \\   
& $ -7.373698$ & 0.018519 &  & $-16.389235$ & 0.026047 \\   
& $ -7.234546$ & 0.092593 &  & $-15.873510$ & 0.001860 \\   
\hline                                                      
m                                                           
& $ -0.252188$ & 0.033333 &  & $ -0.589779$ & 0.020930 \\   
& $ -0.213155$ & 0.092593 &  & $ -0.531971$ & 0.010631 \\   
& $ -0.133040$ & 0.466667 &  & $ -0.502427$ & 0.011628 \\   
& $ -0.073991$ & 0.018519 &  & $ -0.013336$ & 0.146512 \\   
& $  0.189495$ & 0.300000 &  & $ -0.003910$ & 0.499668 \\   
& $  0.308643$ & 0.033333 &  & $  0.013298$ & 0.267442 \\   
& $  0.323117$ & 0.018519 &  & $  0.524366$ & 0.011628 \\   
& $  0.462268$ & 0.037037 &  & $  0.529035$ & 0.010631 \\   
&              &          &  & $  0.563171$ & 0.020930 \\   
\hline                                                      
is                                                          
& $  7.350050$ & 0.018519 &  & $ 15.883012$ & 0.001860 \\   
& $  7.469198$ & 0.166667 &  & $ 16.398737$ & 0.026047 \\   
& $  7.886323$ & 0.092593 &  & $ 16.411175$ & 0.037209 \\   
\hline                                                      
os                                                          
& $ 19.319597$ & 0.148148 &  & $ 25.981267$ & 0.020930 \\   
& $ 19.845140$ & 0.074074 &  & $ 26.042011$ & 0.037674 \\   
&              &          &  & $ 26.557749$ & 0.004186 \\   
\hline
Sum &          & 2.000000 &  &              & 1.255814 \\
\hline
\end{tabular} 
\end{table}

The use of 
$A^*_m\equiv A(1-\exp\{-\tau_m\})$ and 
$\tau^*_m\equiv 1-\exp\{-\tau_m\}$ 
as fit parameters instead of 
$A$ or $A\tau_m$, and $\tau_m$
needs some justification. 
First, 
$A^*_m$
is well determined even in the two limiting cases, 
$\tau_m\ll1$ (with $A$ ill-determined) and 
$\tau_m\gg1$ (with $A\tau_m$ ill-determined).
This is because 
$A^*_m$ 
is a true physical magnitude, i.e. the peak intensity of the main line, provided
that the the hyperfine lines that may compose the main line are blended. 
Second, $\tau^*_m$ is bounded, $0<\tau^*_m<1$,
and the two bounds correspond to the optically thin
and thick limiting cases, 
where the fit is degenerate, and no longer depends on 
$\tau_m$. 
In short, the two fit parameters are well determined in all cases, even in the
optically thin and thick limits. 
The only drawback of the use of 
$\tau^*_m$
as fit parameter is for the case of extremely high opacity in the simultaneous
fit of \nh$(1,1)$ and $(2,2)$, when the intensity of the $(2,2)$ main and 
satellite lines is similar. 
In its present version, \hfsnh\ is not able to deal accurately with 
$\tau_{1m}>16$, but this case is indeed very rare. 
In a survey of low-mass cores in Perseus \citep{Ros08}, none of the sources has
such a high value of optical depth, and in another survey of high-mass clumps in
the Galactic Plane \citep{Svo16} only 1\% of the sources show $\tau_{1m}>16$.
Thus, \hfsnh\ should work for the vast majority of cases.

The fitting procedure ends with a set of four or,
for the \hfsnh\ procedures, six values plus $\tau^*_{2m}$,
of the fit parameters for every velocity component, 
which minimize the fit residual $\chi^2$ of the spectrum (see \S
\ref{sfitting}), 
and an estimation of the uncertainty of the fit parameters (see \S \ref{sconf}),
$\sigma(\Delta V)$,
$\sigma(\vlsr)$,
$\sigma(A^*_m)$, 
$\sigma(\tau^*_m)$, 
and, for the \hfsnh\ procedures,
$\sigma(V_\mathrm{LSR2})$, and
$\sigma(A^*_{2m})$.


\section{Fitting strategy}
\label{sfitting}

The fitting procedure is similar to that used in other fitting problems by 
\citet{Est12} and \citet{Pal14}.
\hfs\ samples the space parameter of dimension $m$ (four or six
times the number of velocity components), defined by the parameters
$p_1,\ldots,p_m$, to find the minimum value of the fit residual $\chi^2$,
\begin{equation}
\chi^2=\sum_{i=1}^N
\left[
\frac{y_i^\mathrm{obs}-y_i^\mathrm{mod}(p_1,\ldots,p_m)}{\sigma_i}
\right]^2,
\end{equation}
where 
$y_i^\mathrm{obs}$ are the observed line intensities, for a total of $N$
spectral channels,
$y_i^\mathrm{mod}(p_1,\ldots,p_m)$ are the model line intensities, depending on
$m$ free parameters, and
$\sigma_i$ are the errors of the observations.

The sampling strategy is based on that used in AGA
(Asexual Genetic Algorithm) \citep{Can09}. The
procedure starts with a number of samples of the parameter space within the
initial search range for each parameter (seeds). For each seed, the residual
$\chi^2$ is computed for a number of samples (descendants) within the search
range centered on the seed value. The best samples, i.e.\ those with the lowest
$\chi^2$, are kept as seeds for the next loop, for which the search ranges are
decreased by a constant factor. The procedure is iterated, and is stopped after
a given number of loops.

Several sampling methods of the $m$-dimensional parameter space are possible,
i.e.\ regular grid, random, Halton or Sobol 
pseudo-random sequences \citep{Hal64, Sob67}. \hfs\ uses a Sobol pseudo-random 
sequence because it
samples the $m$-dimensional parameter space more evenly than a purely random
sequence, even for high values of $m$, and the convergence of the fitting
procedure to the minimum of $\chi^2$ is faster.

\section{Synthetic spectrum}
\label{ssynt}

Let us assume that the transition being fitted has $n_h$ hyperfine lines, and 
for each hyperfine
$j=1,\ldots n_h$, 
$V_\mathrm{hyp}^j$ is the velocity shift with respect to the main line, and
$\tau_\mathrm{hyp}^j$ is the ratio of the optical depth of the hyperfine and
that of the main line, $\tau_m$.
Let $n_c$ be the number of velocity components fitted, and for each velocity 
component 
$i=1,\ldots n_c$, 
$\Delta V^i$, 
$\vlsr^i$, 
$A^i$, and
$\tau_m^i$ 
are the values of the fit or derived line parameters (see \S \ref{sderived}).
 
For each velocity channel
$k=1,\ldots N$, with central velocity $V_k$ and channel width $\Delta 
V_\mathrm{ch}$,
the contribution of the velocity component $i$ to the optical depth is 
calculated as
\begin{equation}
  \tau_k^i= \sum_{j=1}^{n_h} 
\frac{\tau_m^i \tau_\mathrm{hyp}^j}{\Delta V_\mathrm{ch}}\,G,
\end{equation}
where $G$ is the integral over the velocity range of the channel,
\begin{equation}
G=\int_{V_k-\Delta V_\mathrm{ch}/2}^{V_k+\Delta V_\mathrm{ch}/2}
e^{-4\ln2\left[\left(v-\vlsr^i-V_\mathrm{hyp}^j\right)/\Delta V^i\right]^2}\,dv.
\end{equation}
$G$ is evaluated by means of the error function 
$\mathrm{erf}\,(x)=2/\!\sqrt{\pi}\int_0^x \exp(-t^2) dt$, as
\begin{equation}\label{diferf}
G=  \frac{\sqrt{\pi}}{4\sqrt{\ln{2}}} \Delta V^i 
[\mathrm{erf}\,(x^+)-\mathrm{erf}\,(x^-)],
\end{equation}
where 
\begin{equation}
  x^\pm= 2\sqrt{\ln2}\, \displaystyle\frac{V_k\pm\Delta 
V_\mathrm{ch}/2-\vlsr^i-V_\mathrm{hyp}^j}{\Delta V^i}.
\end{equation}
If the difference between $x^+$ and $x^-$ is too low, within less than 1 part in
$10^4$, Equation \ref{diferf} can produce a large roundoff error, and  $G$ is
evaluated as
\begin{equation}
G=  
\frac{\Delta V^i}{2\sqrt{\ln{2}}}\, (x^+-x^-)\, e^{-\left[(x^++x^-)/2\right]^2}.
\end{equation}
Finally, the intensity of the channel $k$ of the synthetic spectrum is 
calculated as the sum of intensities of the different velocity components,
\begin{equation}
  T_k=\sum_{i=1}^{n_c} A^i \left(1-e^{-\tau_k^i}\right).
\end{equation}

\section{Derived line parameters}
\label{sderived}

From the values of the fit parameters, the following derived line parameters are
calculated, which are necessary for calculating the synthetic spectra and the
estimation of parameters with physical interest:

\paragraph{$A$, amplitude.} 
The amplitude is 
\begin{equation}
A=f[J_\nu(\Tex)-J_\nu(\Tbg)], 
\end{equation}
where 
$f$ is the beam filling factor, 
$\Tex$ the excitation temperature,  
$\Tbg$ the background temperature, and
\begin{equation}
J_\nu(T)=\frac{h\nu/k}{e^{h\nu/kT}-1}, 
\end{equation}
is the Planck-corrected temperature. $A$ is calculated from
\begin{equation}
A = \frac{A^*_m}{\tau^*_m}.
\end{equation}
In the case of \hfsnh, the excitation temperature $\Tex$ and the filling factor
$f$ are assumed to be the same for the $(1,1)$ and $(2,2)$ lines. Since the
frequencies of both transitions are very close (see Table \ref{tnu}), 
these assumptions imply that the amplitude $A$ is the
same for both lines, within less than 1 in $10^4$.

\paragraph{$\tau_m$, optical depth of the main line.} 
Calculated from
\begin{equation}
\tau_m = -\ln(1-\tau^*_m).
\end{equation}
Special care has to be taken when $\tau_m\simeq\tau^*_m\ll1$, since the
last expression involves the difference of 1 and a number near 1. In this case,
a good approximation is the Taylor expansion
\begin{equation}
\tau_m\simeq
\tau^*_m+
\frac{1}{2}{\tau^*_m}^2+
\frac{1}{3}{\tau^*_m}^3.
\end{equation}
In the case of \hfsnh, $\tau_{1m}$ and $\tau_{2m}$ are derived in the same way.

\paragraph{$A\tau_m$, amplitude times the main line optical depth.}
Calculated from
\begin{equation}
A\tau_m = A^*_m \frac{\tau_m}{\tau^*_m}.
\end{equation}
Note that for $\tau^*_m\ll1$, $A\tau_m\simeq A^*_m$.
In the case of \hfsnh, $A\tau_{1m}$ and $A\tau_{2m}$ are derived in the same 
way.

\section{\nh\ derived physical parameters}
\label{sphyspar}

In addition to the derived line parameters, \hfsnh\ derives physical parameters 
using the standard analysis of NH$_3$
$(1,1)$ and $(2,2)$ observations, which assumes that the region observed 
is homogeneous
along the line of sight. The physical parameters derived are 
the excitation temperature $\Tex$, 
the NH$_3$ $(1,1)$ and $(2,2)$ beam-averaged column densities $fN(1,1)$ and
$fN(2,2)$, 
the rotational temperature $T_\mathrm{rot}$,  the NH$_3$
beam-averaged column density $fN(\mathrm{NH}_3)$, and 
the kinetic temperature $T_k$.

\paragraph{Excitation temperature.}
The excitation temperature is obtained from the amplitude $A$, 
\begin{equation}
\Tex= 
\frac{\Tnu}{\ln 
\left[1+\dfrac{\Tnu}{A/f+J_\nu(\Tbg)}\right]},
\end{equation}
where $\Tnu=h\nu/k$ is the frequency in temperature units.
The frequency of the $(1,1)$ transition is used, but since the frequency of the
$(2,2)$ transition is very close (see Table \ref{tnu}), the result does not
depend on which of the two frequencies is used. 
For values of $\Tex\gg\Tbg>\Tnu$, this expression simplifies to
\begin{equation}
\Tex\simeq A/f. 
\end{equation}
The value of the excitation temperature depends on the value assumed for the
filling factor $f$.
The usual assumption is that the filling factor $f=1$. The value obtained with
this assumption is a lower limit for the value of $\Tex$. 
On the contrary, if we assume that $f\ll1$, $\Tex\rightarrow\infty$.

\paragraph{NH$_3$ $(1,1)$ and $(2,2)$ column densities.}
The column density of the $(J,K)=(1,1)$ and $(2,2)$ levels 
(i.e.\ the sum of column densities of the two inversion levels of the
corresponding $(J,K)$ rotational level)
can be given as \citep{Ang95, Est97}
\begin{equation}
N(J,K)= \sqrt{\frac{\pi}{4\ln2}}\,
\frac{8\pi{\nu_{jk}}^3}{c^3 A_{jk}}\, 
R_m\,
\frac{\exp(\Tnu/\Tex)+1}{\exp(\Tnu/\Tex)-1}\,
\tau_m \Delta V,
\end{equation}
where
$A_{jk}$ is the Einstein coefficient of the inversion transition of the
rotational level $(J,K)$,
$R_m=\tau_\mathrm{tot}/\tau_m$ is the ratio of total and main line
optical depths of the inversion transition, and 
$\Tnu=h\nu_{JK}/k$ is the frequency of the inversion transition in temperature
units (see Table \ref{tnu}).

\begin{table*}[htb]
\centering
\caption{\label{tnu}
Values of the NH$_3$ $(1,1)$ and $(2,2)$ inversion transition 
frequencies \citep{Kuk67}, 
spontaneous emission Einstein coefficients \citep{Oso09}, 
ratios of total to main line optical depths \citep{Man15}, and 
$B(J,K)$ and $C(J,K)$ coefficients of \citet{Ang95} recalculated with
the improved values of the constants in this table.}
\begin{tabular}{lcccccc}
\hline\hline
         & $\nu_{jk}$ & $T_\nu=h\nu_{jk}/k$  & $A_{jk}$   & $R_m=$ \\
$(J,K)$  & (GHz)  & (K) & ($10^{-7}$ s$^{-1})$ & $\tau_\mathrm{tot}/\tau_m$ & 
$B(J,K)$ & $C(J,K)$ \\ 
\hline
$(1,1)$  & $23.69450$   & $1.137157$ & $1.66838$ & $2.000000$  & 
$1.58339\times10^{13}$ & $2.78482\times10^{13}$ \\
$(2,2)$  & $23.72263$   & $1.138507$ & $2.23246$ & $1.255814$  & 
$7.45665\times10^{12}$ & $1.30990\times10^{13}$ \\
\hline
\end{tabular}
\end{table*}

The last expression depends on the value of the filling factor $f$ assumed to
derive $\Tex$. The explicit dependence on $f$ is
\begin{equation}
\frac{\exp(\Tnu/\Tex)+1}{\exp(\Tnu/\Tex)-1}=
\frac{2}{f\Tnu}(A+f[J_\nu(\Tbg)+\Tnu/2]),
\end{equation}
so that the beam-averaged column density can be expressed as
\begin{eqnarray}
fN(J,K)&=&
\sqrt{\frac{\pi}{4\ln2}}\,
\frac{16\pi k{\nu_{jk}}^2}{h c^3 A_{jk}}\, 
R_m\times   \nonumber \\
&\times& \left(A\tau_m+f\left[J_\nu(\Tbg)+\frac{\Tnu}{2}\right]\tau_m\right)\, 
\Delta V.
\end{eqnarray}
The maximum value of $fN(J,K)$ is obtained for $f=1$ (the usual assumption to
derive $\Tex$), while the minimum value is obtained for $f\ll1$. In the latter
case (or for $A\simeq\Tex\gg\Tbg>\Tnu$), the expression simplifies to
\begin{equation}
fN(J,K)= \sqrt{\frac{\pi}{4\ln2}}\,
\frac{16\pi k{\nu_{jk}}^2}{h c^3 A_{jk}}\, R_m\, A\tau_m\, \Delta V.
\end{equation}
The values of the constants appearing in these equations
are given in Table \ref{tnu}.

In practical units the two equations become \citep{Ang95}
\begin{equation}
\left[\frac{N(J,K)}{\mathrm{cm^2}}\right] = B(J,K)\,
\frac{\exp(\Tnu/\Tex)+1}{\exp(\Tnu/\Tex)-1}\,
\tau_m 
\left[\frac{\Delta V}{\mathrm{km\ s^{-1}}}\right] 
\end{equation}
and, for $f\ll1$, or $\Tex\gg\Tbg>\Tnu$,
\begin{equation}
\left[\frac{fN(J,K)}{\mathrm{cm^2}}\right]= C(J,K)\,
A\tau_m 
\left[\frac{\Delta V}{\mathrm{km\ s^{-1}}}\right],
\end{equation}
The values of the constants $B(J,K)$ and $C(J,K)$ 
for the $(1,1)$ and $(2,2)$ transitions are given in Table 2.
The expression equivalent to these two equations, 
with an explicit dependence on $f$, is,
\begin{equation}
\left[\frac{fN(J,K)}{\mathrm{cm^2}}\right] = C(J,K)\,
\left(A\tau_m+f\left[J_\nu(\Tbg)+\frac{\Tnu}{2}\right]\tau_m\right) 
\left[\frac{\Delta V}{\mathrm{km\ s^{-1}}}\right].
\end{equation}

\paragraph{Rotational temperature.}
The rotational temperature is obtained from the ratio of $(1,1)$ and
$(2,2)$ column densities,
\begin{equation}
\Trot= \frac{(E_{22}-E_{11})/k}
{\ln\left(\dfrac{g_{22}}{g_{11}}\dfrac{N(1,1)}{N(2,2)}\right)},
\end{equation}
where $E_{11}$, $E_{22}$, and $g_{11}$, $g_{22}$, are the energies and
degeneracies of the corresponding levels. 
In practical units the equation becomes (see Table \ref{ttrot}),
\begin{equation}
\left[\frac{\Trot}{\mathrm{K}}\right]= 
\frac{40.99}{\ln\left(\dfrac{5}{3}\,\dfrac{N(1,1)}{N(2,2)}\right)}.
\end{equation}

\begin{table}[htb]
\centering
\caption{\label{ttrot} 
Degeneracies and energies above the $(1,1)$ level of the lower metastables 
levels of NH$_3$ \citep{Poy75, Man15}. Note that the values of the energies
reported by these authors are slightly different from those given in
\citet{Ho83}.}
\begin{tabular}{ccc}
\hline\hline
        &          & $(E_{JK}-E_{11})/k$ \\
$(J,K)$ & $g_{JK}$ & (K) \\
\hline
$(0,0)$ & $1/3$  & $          -22.64$ \\   
$(1,1)$ & $1$    & $\phantom{-2}0.00$ \\   
$(2,2)$ & $5/3$  & $\phantom{-}40.99$ \\   
$(3,3)$ & $14/3$ & $\phantom{-}99.76$ \\   
\hline
\end{tabular}
\end{table}

\paragraph{NH$_3$ column density.}
The ammonia total column density is usually estimated assuming that the 
population ratios between pairs of levels are given by the same rotational
temperature (CTEX approximation),  and that only the metastable levels $J=K$, up
to $(J,K)=(3,3)$, are populated. 
While this assumption is reasonable for low-mass dense cores with moderate
temperatures, it is not appropriate for hot cores, where \nh\ inversion 
transitions $(6,6)$ to $(14,14)$ are detected \citep[see for instance][]{God15}.

In addition, since the \nh$(1,1)$ and $(2,2)$ levels correspond to para-\nh, no
information on the ortho-\nh\ is obtained from the $(1,1)$ and $(2,2)$ spectra.
Thus, an ortho-to-para ratio has to be assumed. The usual assumption is to take an
ortho-to-para ratio of 1, although a value of $\sim0.7$ has been observed in
some star-forming regions \citep[see for instance the discussion in][]{Fau13}. 
We will assume an ortho-to-para ratio of 1.

With the former assumptions,
\begin{equation}
\Nnht= N(1,1)\,Q,
\end{equation}
with the partition function $Q$ given by
\begin{equation}
Q=
\sum_{J,K=0}^{3} \frac{g_{JK}}{g_{11}} e^{(E_{11}-E_{JK})/k\Trot}.
\end{equation}
In practical units (see Table 3),
\begin{eqnarray}
\Nnht&=& N(1,1) \left[
\frac{1}{3}  e^{ 22.64/\Trot}+
1+                              \right.\nonumber \\ &+&\left.
\frac{5}{3}  e^{-40.99/\Trot}+  
\frac{14}{3} e^{-99.76/\Trot}
\right].
\end{eqnarray}

\paragraph{Kinetic temperature.}

\begin{figure}[thb]
\centering
\plotone{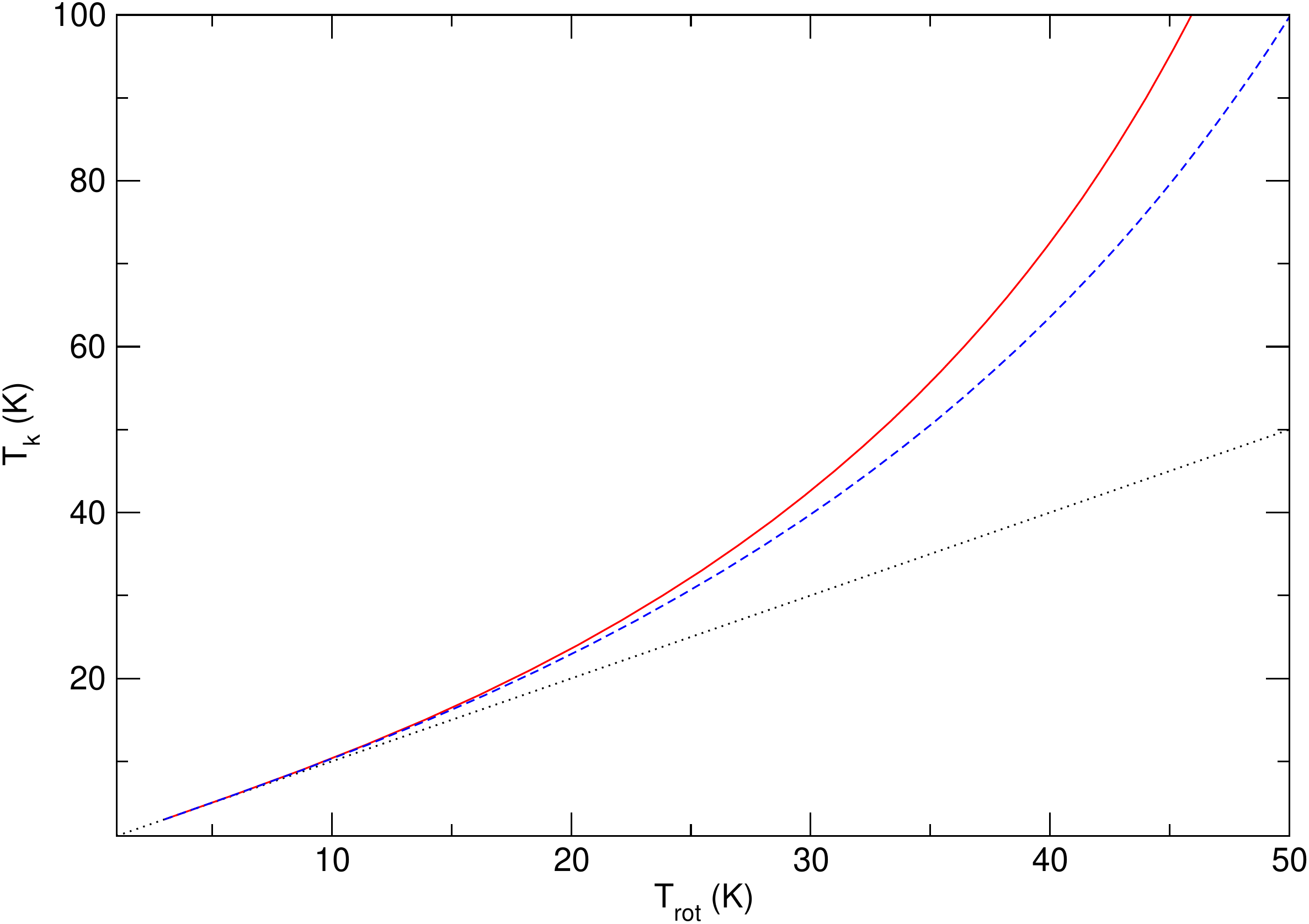}
\caption{\label{ftk}
$T_k$ as a function of $\Trot$. 
Red continuum line: \citet{Mar09} and present work;
blue dashed line: \citet{Swi05}, \citet{Ros08}, \citet{Man15}; 
black dotted line: $T_k=\Trot$.
}
\end{figure}

The kinetic temperature can be taken to be equal to the rotational
temperature $\Trot$, but a better estimation can be given taking into account
collisional transitions to other levels. 
A 3-level approximation considering only the rotational levels  $(1,1)$,
$(2,2)$, and $(2,1)$ can be shown to be \citep{Wal83, Dan88, Man15}
\begin{equation}
\Trot= \dfrac{T_k}{1+\dfrac{T_k}{T_0}
\ln\left(1+\dfrac{C(22\to21)}{C(22\to11}\right)},
\end{equation}
where $T_0=(E_{22}-E_{11})/k$, and
$C(22\to21)$ and $C(22\to11)$ 
are the collisional excitation and desexcitation rates between the
corresponding levels.
Calculations involving more rotational levels were performed by \citet{Dan88},
and more recently by \citet{Mar09} using improved values of the collisional
rates. 
A good approximation to the results of \citet{Mar09} can be given
as a slightly modified version of the widely used expression  
\citep[see Fig.\ \ref{ftk}]{Swi05, Ros08, Man15}, which is
\begin{equation}
\Trot= \dfrac{T_k}{1+\dfrac{T_k}{T_0}
\ln\left[1+0.73\,e^{-(E_{21}-E_{22})/kT_k}\right]},
\end{equation}
with $T_0= 40.99$ K, $(E_{21}-E_{22})/k= 16.26$ K, and the numerical value
$0.73$ was determined to fit the results shown in Fig.\ 5 of \citet{Mar09}. Note
that this relation implies that $\Trot$ is always below a value 
$\Trot=40.99/\ln1.73=74.8$~K (see Fig.1). 
Given a value of $\Trot$, 
the implicit equation must be solved to find $T_k$.
A possible iterative algorithm to solve the equation is
\begin{equation}
T_k^{(n+1)}= \Trot 
\left[1+
\dfrac{T_k^{(n)}}{40.99}\ln\left(1+0.73\,e^{-16.26/T_k^{(n)}}\right)\right],
\end{equation}
starting with $T_k^{(0)}=\Trot$.

\section{Error estimation of the fit parameters}
\label{sconf}

\subsection{Parameter-space confidence region}

Let us assume that the best fit to the observed data
is obtained for values $p^0_k$ of the parameters, for which
the residual $\chi^2$ is minimum,
\begin{equation} 
\chi^2_\mathrm{min}=\sum_{i=1}^N
\left[
\frac{y_i^\mathrm{obs}-y_i^\mathrm{mod}(p^0_1,\ldots,p^0_m)}{\sigma_i}
\right]^2,
\end{equation}
where $\sigma_i$ is the error of $y_i^\mathrm{obs}$.
The uncertainty, $\sigma(p_k)$, in the values derived for the parameters $p_k$
can be estimated as the projection of the confidence region of the
$m$-dimensional space parameter for which $\chi^2$ does not exceed the minimum
value by an amount $\Delta(m,\alpha)$, where $\alpha$
is the significance level ($0<\alpha<1$). 
Following \citet{Avn76} and \citet{Wal03}, the probability 
\begin{equation}
\mathrm{Prob}\,[\chi^2-\chi^2_\mathrm{min}\le\Delta(m,\alpha)]=\alpha,
\end{equation}
is that of a chi-square distribution with $m$ degrees of freedom. 
Thus, $\Delta(m,\alpha)$ is the increment of $\chi^2$ such
that if the observation is repeated a large number of times, a fraction $\alpha$
of times the values of the parameters fitted will be inside the confidence
region, i.e.\  in the interval $p_k\pm\sigma(p_k)$. 
The values of $\Delta(m,\alpha)$ for significance levels equivalents to 1, 2, 
and 3 sigmas for a Gaussian error distribution, and different values of $m$ are 
shown in Table \ref{tdelta}.

\begin{table}[htb]
\centering
\caption{\label{tdelta}
Values of $\Delta(m, \alpha)$ for calculating the parameter uncertainties, where
$m$ is the number of parameters fitted simultaneously, and $\alpha$ is the
significance level, given in percent and in the equivalent number of sigmas for
a Gaussian error distribution. The values shown are for values of $m$  
multiple of 4 or 6, used by \hfs.
} 
\begin{tabular}{rrrr}
\hline\hline
    &\multicolumn{3}{c}{$\alpha$}\\
\cline{2-4}
    &      68.27\% &      95.45\% &     99.73\% \\
$m$ & (1 $\sigma$) & (2 $\sigma$) & (3 $\sigma$)\\
\hline
 4 &  4.72 &   9.72 &  16.25  \\
 6 &  7.04 &  12.85 &  20.06  \\
 8 &  9.30 &  15.79 &  23.57  \\
12 & 13.74 &  21.35 &  30.09  \\
16 & 18.11 &  26.65 &  36.21  \\
18 & 20.28 &  29.24 &  39.17  \\
20 & 22.44 &  31.80 &  42.08  \\
24 & 26.73 &  36.83 &  47.76  \\
28 & 31.00 &  41.78 &  53.31  \\
30 & 33.12 &  44.22 &  56.04  \\
32 & 35.24 &  46.66 &  58.75  \\
36 & 39.48 &  51.48 &  64.10  \\
42 & 45.80 &  58.64 &  71.99  \\
48 & 52.11 &  65.72 &  79.75  \\
54 & 58.39 &  72.72 &  87.41  \\
\hline
\end{tabular}
\end{table}

Assuming that the model fits well the observations, i.e.\
$\chi^2_\mathrm{min}\simeq N-m$, the condition 
\begin{equation}
\chi^2=\chi^2_\mathrm{min}+\Delta(m,\alpha),
\end{equation}
can be written as
\begin{equation}
\frac{\chi^2}{\chi^2_\mathrm{min}}\simeq1+\frac{\Delta(m,\alpha)}{N-m}.
\end{equation}
This expression is useful since it can be given in terms of
the weighted rms fit residual, $\sigma$,
\begin{equation}
\sigma^2=
\frac
{\sum_{i=1}^N\left[(y_i^\mathrm{obs}-y_i^\mathrm{mod})/\sigma_i\right]^2}
{\sum_{i=1}^N 1/\sigma_i^2}=
\frac{\chi^2}{\sum_{i=1}^N 1/\sigma_i^2},
\end{equation}
for which we obtain that the confidence region is given by the parameter values
that increase the rms fit residual to
\begin{equation}
\sigma\simeq\sigma_\mathrm{min}
\sqrt{1+\frac{\Delta(m,\alpha)}{N-m}}.
\end{equation}
This last expression can be used even when the errors of the observations are
unknown.

\subsection{Modeling the fit residual $\chi^2$}

\subsubsection{Quadratic approximation of the fit residual}

Let us assume that around its minimum value, the residual $\chi^2$ can be
approximated by a quadratic function,
\begin{equation}\label{eqchi2}
\chi^2 \simeq \chi^2_\mathrm{min} +
\sum_{i, j= 1}^m a_{ij} \, x_i \, x_j+ 
2 \sum_{i= 1}^m b_i \, x_i,
\end{equation}
where $x_j= p_j-p^0_j$ are the increment of the parameter values from their
best-fit values. 
The confidence region of the $m$-dimensional parameter space 
will be the region inside the surface
\begin{equation}
\sum_{i, j= 1}^m a_{ij} \, x_i \, x_j+ 
2 \sum_{i= 1}^m b_i \, x_i = \Delta(m, \alpha),
\end{equation}
which is the general equation of a $m$-dimensional quadric \citep{McC11}.
This equation depends on a total of $m(m+3)/2$ coefficients: 
$m(m+1)/2$ symmetric $a_{ij}$ coefficients ($a_{ij}=a_{ji}$), and
$m$ coefficients $b_i$.

In array form, the quadric equation can be expressed as
\begin{equation}
\left(\begin{array}{cccc}
x_1 & \ldots  & x_m & 1
\end{array}\right)
\left(\begin{array}{cccc}
a_{11} & \ldots & a_{1m} & b_1    \\
\vdots & \ddots & \vdots & \vdots \\
a_{m1} & \ldots & a_{mm} & b_m    \\
b_1    & \ldots & b_m    & c
\end{array}\right)
\left(\begin{array}{c}
x_1 \\ \vdots \\ x_m \\ 1
\end{array}\right)
= 0,
\end{equation}
with $c=-\Delta(m, \alpha)$,
or, with the obvious definitions for arrays $\mathbf{A}$, $\mathbf{B}$, and
$\mathbf{X}$,
\begin{equation}
\left(\begin{array}{cc}
\mathbf{X}^t & 1
\end{array}\right)
\left(\begin{array}{cc}
\mathbf{A} & \mathbf{B} \\
\mathbf{B}^t & c          \\
\end{array}\right)
\left(\begin{array}{c}
\mathbf{X} \\ 1
\end{array}\right)
= 0.
\end{equation}

In order to estimate the uncertainties of the fit parameters, $\sigma(p_k)$,
we have to calculate the projections of the quadric onto each axis $k$.
If the fit residual is well behaved, we may expect that the quadric is an 
ellipsoid, i.e.\ its
projections onto the plane defined by any pair of parameters is an ellipse, and
the ellipsoid has finite projections onto any axis.

\subsubsection{Projections of an ellipsoid}

The projections of the ellipsoid onto each coordinate axis can be found as the
intersections of the hyperplane perpendicular to the axis, tangent to the
ellipsoid. The equation of the tangent hyperplane at a point 
$(x^0_1, \ldots, x^0_m)$ of
the ellipsoid is given by \citep{McC11}
\begin{equation}
\left(\begin{array}{cccc}x^0_1 & \ldots  & x^0_m & 1\end{array}\right)
\left(\begin{array}{cccc}
a_{11} & \ldots & a_{1m} & b_1    \\
\vdots & \ddots & \vdots & \vdots \\
a_{m1} & \ldots & a_{mm} & b_m    \\
b_1    & \ldots & b_m    & c
\end{array}\right)
\left(\begin{array}{c}x_1 \\ \vdots \\ x_m \\ 1\end{array}\right)
= 0,
\end{equation}
or
\begin{equation}
\sum_{i, j= 1}^m a_{ij} \, x^0_i \, x_j + 
\sum_{j= 1}^m b_j \, x_j + \sum_{i= 1}^m b_i \, x^0_i +
c= 0.
\end{equation}
The equation of an hyperplane perpendicular to the $k$ axis is
$x_k=$ constant, 
so that  the equation of the tangent hyperplane is
\begin{equation}\label{eqtangent}
\left[\sum_{i= 1}^m a_{ik} \, x^0_i + b_k\right] x_k+ 
\sum_{i= 1}^m b_i \, x^0_i + c = 0,
\end{equation}
and  the coefficients of $x_j$ for $j\neq k$ have to be zero,
\begin{equation}\label{eqcoef}
\sum_{i= 1}^m a_{ij} \, x^0_i + b_j=0 \qquad (j\neq k).
\end{equation}
Eqs.\ \ref{eqcoef} form a system of $m-1$ linear equations with $m$
unknowns, the coordinates of the tangent point. We can consider that 
$x^0_j$ ($j\neq k$) are the $m-1$
unknowns  of the system, which can be derived as a function of $x^0_k$.
The system of equations can be written with arrays of size $m$,
\begin{equation}
\mathbf{A}_k \, \mathbf{X}= \mathbf{Z}_k \, x^0_k + \mathbf{C}_k
\end{equation}
where $\mathbf{A}_k$ is the array $\mathbf{A}$ with zeros in the $k$ row and
column, $a_{ik}=a_{ki}=0$, and 1 in the diagonal term, $a_{kk}=1$, and 
$\mathbf{Z}_k$ and $\mathbf{B}_k$ have $1$ and 0 respectively in the $k$ row, 
\begin{eqnarray}\label{eqakzk}
\mathbf{A}_k= \left(\begin{array}{ccccc}
a_{11} & \ldots & 0      & \ldots & a_{1m} \\
\vdots &        & \vdots &        & \vdots \\
0      & \ldots & 1      & \ldots & 0      \\
\vdots &        & \vdots &        & \vdots \\
a_{m1} & \ldots & 0      & \ldots & a_{mm}
\end{array}\right), 
\quad
\nonumber\\
\mathbf{Z}_k= \left(\begin{array}{c}
-a_{1k} \\
\vdots \\
1     \\
\vdots \\
-a_{mk}
\end{array}\right), \quad
\mathbf{C}_k= \left(\begin{array}{c}
-b_1    \\
\vdots \\
0      \\
\vdots \\
-b_m
\end{array}\right).
\end{eqnarray}
From the system we can derive the solution giving $x^0_j$ in terms of $x^0_k$, 
\begin{equation}
\mathbf{X}= \mathbf{D}_k \, x^0_k + \mathbf{E}_k, \quad\mathrm{with}\quad
\mathbf{D}_k= \mathbf{A}_k^{-1} \mathbf{Z}_k, \quad 
\mathbf{E}_k= \mathbf{A}_k^{-1} \mathbf{C}_k,
\end{equation}
which can be expressed as
\begin{equation}\label{eqsystem}
x^0_j= d_{jk} \, x^0_k + e_{jk} \qquad (j= 1, \ldots, m),
\end{equation}
where $d_{kk}=1$ and $e_{kk}=0$. 

The tangent point must fulfill the tangent hyperplane equation, Eq.\
\ref{eqtangent}. By substitution of Eq.\ \ref{eqsystem} in the tangent
hyperplane equation,  and setting $x_k = x^0_k$, we get a second degree equation
in $x^0_k$,
\begin{eqnarray}\label{eq2nd}
\left[\sum_{i= 1}^m a_{ki} \, d_{ik}\right] \left(x^0_k\right)^2 &+&
\left[\sum_{i= 1}^m\left(a_{ki} \, e_{ik} + b_i \,d _{ik}\right) + b_k\right]  
x^0_k +
\nonumber\\ &+&
\left[\sum_{i= 1}^m b_i \, e_{ik} + c\right] = 0.
\end{eqnarray}
The two solutions of the equation provide the two $k$-coordinates of the
projections of the ellipsoid onto the $k$ axis.

\subsubsection{Case of a centered ellipsoid}

Since we are assuming that we know that the minimum residual $\chi^2$ is well 
determined, the quadratic function of Eq.\ \ref{eqquadric} must have a minimum at the 
origin, and the linear terms in the variables vanish because the partial 
derivatives at the origin must be zero. This means that we can assume that the 
quadric is centered, and its equation becomes
\begin{equation}\label{eqquadric}
\sum_{i, j= 1}^m a_{ij} \, x_i \, x_j + c= 0.
\end{equation}
In this case, the equation of the tangent plane is simpler, since 
$\mathbf{B}=0$,
$\mathbf{C}_k=0$, and
$\mathbf{E}_k=0$.
The second degree equation (Eq.\ \ref{eq2nd}) becomes
\begin{equation}\label{eqx02}
\left[\sum_{i= 1}^m a_{ki} \, d_{ik}\right] \left(x^0_k\right)^2 + c = 0,
\end{equation}
and the projections are symmetric, $\pm x^0_k$.

\subsubsection{Coefficients of the ellipsoid}

The centered ellipsoid depends on 
$m(m+1)/2$ symmetric $a_{ij}$ coefficients.
Let us examine a sufficient number of constraints to derive these coefficients.

\paragraph{Diagonal coefficients $a_{ii}$.}

For each parameter $i$ we can obtain constraints from the values of the residual
$\chi^2$ increment for different increments $x_i$ of the parameter $p_i$, while
keeping the rest of parameters constant, 
\begin{equation}
\chi^2(p^0_1, \ldots, p^0_i+x_i, \ldots, p^0_m)=
\chi^2_\mathrm{min}+\Delta_i.
\end{equation}
For each value of the increments of the parameter and the residual
$\chi^2$ increment, $x^n_i$, $\Delta^n_i$, we have a constraint on $a_{ii}$,
\begin{equation}\label{eqakk}
a_{ii} \, (x^n_i)^2 \simeq \Delta^n_i. 
\end{equation}
Although a single value is enough to determine $a_{ii}$, at least two are
recommended, above and below the best-fit value, i.e.\ with $x_i>0$ and $x_i<0$.
In general, the best approximation for $a_{ii}$ (so that the sum of the squares
of $[\Delta^n_i-a_{ii} \, (x^n_i)^2]$ is minimum) is given by
\begin{equation}\label{eqakkval}
a_{ii}= \displaystyle
\frac{\sum_n \Delta^n_i (x^n_i)^2}{\sum_n (x^n_i)^4}
\end{equation}
where the sums are for all increments evaluated. 
Note that, provided that the $\Delta^n_i$ are positive, the diagonal term
$a_{ii}$ will always be positive. For a good characterization of the behavior of
$\chi^2$, at least two values of $\Delta^n_i$ should be close to 
$\Delta(m,\alpha)$.

The intersections of the ellipsoid with the coordinate axis $i$ are
$\pm({\Delta(m,\alpha)/a_{ii}})^{1/2}$. 
For the case of statistically independent parameters, these intersections will
coincide with the projections of the ellipsoid, since  the coordinate axes are
the principal axes of the ellipsoid and the cross-terms of the ellipsoid
vanish.   However, in general, there will be some dependence among the
parameters, and the cross terms will not be zero.

\paragraph{Cross-coefficients $a_{ij}$ ($i\neq j$).}

The constraints to derive the cross terms $a_{ij}$ can be
obtained from the value of the residual $\chi^2$ for the simultaneous increment
of the two parameters $i$ and $j$, while
keeping the rest of parameters constant, 
\begin{equation}
\chi^2(p^0_1, \ldots, p^0_i+x_i, \ldots, p^0_j+x_j, \ldots, p^0_m)=
\chi^2_\mathrm{min}+\Delta_{ij}
\end{equation}
For each pair of increments of the parameters and the residual $\chi^2$
increment, $x^n_i$, $x^n_j$, $\Delta^n_{ij}$, we have a constraint on $a_{ij}$, 
\begin{equation}
a_{ii}(x^n_i)^2 + 2a_{ij} x^n_i x^n_j + a_{jj}(x^n_j)^2 \simeq \Delta^n_{ij}.
\end{equation}
Taking into account Eq.\ \ref{eqakk}, and defining 
$\delta^n_{ij}\equiv\Delta^n_{ij}-\Delta^n_i-\Delta^n_j$, 
we have
\begin{equation}
2a_{ij}x^n_ix^n_j \simeq \delta^n_{ij}
\end{equation}
Although a single pair of values $x_i$, $x_j$, is enough to determine $a_{ij}$,
at least four are recommended, above and below the best-fit value for each
parameter, i.e.\ with the four combinations of
$x_i>0$, $x_i<0$, $x_j>0$, and $x_j<0$. 
In general, the best approximation for $a_{ij}$ (in the same sense as for the
diagonal term) is given by
\begin{equation}\label{eqaij}
a_{ij}= \frac{1}{2}\,
\frac{\sum_n \delta^n_{ij} \, x^n_i x^n_j}{\sum_n (x^n_i x^n_j)^2}
\end{equation}
where the sums are for all the pairs of increments evaluated.

\subsubsection{Practical case}

A practical implementation of the procedure is as follows.
Let us call
\begin{equation}
f(x_1,\ldots,x_m)= \chi^2(p^0_1+x_1, \ldots, p^0_m+x_m)-\chi^2_\mathrm{min}.
\end{equation}
\begin{enumerate}

\item
For each parameter $p_i$ we estimate a positive and a negative increment,
$x^+_i$ and $x^-_i$, such that the increment of the residual $\chi^2$ is close
to $\Delta(m,\alpha)$,
\begin{equation}
\begin{array}{l}
f(0, \ldots, x^+_i, \ldots, 0)= \Delta^+_i \simeq \Delta(m,\alpha),\\
f(0, \ldots, x^-_i, \ldots, 0)= \Delta^-_i \simeq \Delta(m,\alpha).
\end{array}
\end{equation}
The diagonal coefficient $a_{ii}$ (Eq.\ \ref{eqakk}) is given by
\begin{equation}
a_{ii}= \frac{\Delta^+_i (x^+_i)^2+\Delta^-_i (x^-_i)^2}{(x^+_i)^4+(x^-_i)^4}.
\end{equation}

\item
For each pair of parameters $p_i$, $p_j$ ($i\neq j$) we calculate
\begin{eqnarray}
&\delta^{++}_{ij}= f(0,\ldots,x^+_i,\ldots, 
x^+_j,\ldots,0)-\Delta^+_i-\Delta^+_j, \nonumber \\
&\delta^{+-}_{ij}= f(0,\ldots,x^+_i,\ldots, 
x^-_j,\ldots,0)-\Delta^+_i-\Delta^-_j, \nonumber \\
&\delta^{-+}_{ij}= f(0,\ldots,x^-_i,\ldots, 
x^+_j,\ldots,0)-\Delta^-_i-\Delta^+_j,           \\
&\delta^{--}_{ij}= f(0,\ldots,x^-_i,\ldots, 
x^-_j,\ldots,0)-\Delta^-_i-\Delta^-_j. \nonumber
\end{eqnarray}
The cross-coefficient $a_{ij}$ (Eq.\ \ref{eqaij}) is given by
\begin{equation}
a_{ij}= \frac{1}{2}\,
\frac{
\delta^{++}_{ij} \, x^+_i x^+_j+
\delta^{+-}_{ij} \, x^+_i x^-_j+
\delta^{-+}_{ij} \, x^-_i x^+_j+
\delta^{--}_{ij} \, x^-_i x^-_j}
{(x^+_i x^+_j)^2+(x^+_i x^-_j)^2+(x^-_i x^+_j)^2+(x^-_i x^-_j)^2}.
\end{equation}

\item
For each parameter $k$ we construct the arrays $\mathbf{A}_k$, $\mathbf{Z}_k$
(Eq.\ \ref{eqakzk}), and solve the system of linear equations
$\mathbf{A}_k \mathbf{D}_k= \mathbf{Z}_k$.
Finally, we calculate the quadratic coefficient, 
$\sum_{i= 1}^m a_{ki} \, d_{ik}$ 
(Eq.\ \ref{eqx02}), and the projection 
$x^0_k=[\Delta(m,\alpha)/\sum_{i= 1}^m a_{ki} \, d_{ik}]^{1/2}$.

\end{enumerate}
In some cases, the quadratic approximation of $\chi^2$ is not good enough, and
the determination of the projections can fail: for some parameter $p_k$ the
array  $\mathbf{A}_k$ may have null determinant, or the quadratic coefficient of
Eq.\ \ref{eqx02} may be negative. In these cases, a rough estimation of the
uncertainty in $p_k$ can still be given as the  intersection with the $k$ axis,
$[\Delta(m, \alpha)/a_{kk}]^{1/2}$.

\section{Error estimation of the derived parameters}
\label{snum}

All the derived parameters depend on $m$ fit parameters
(four or six times the number of velocity component). 
Let us call 
\begin{equation}
p^0_k, \quad k=1,\dots, m
\end{equation}
the values of the fit parameters, and $\sigma(p_k)$ their errors, found from
the  increase in the fit residual $\chi^2$ (see \S \ref{sconf}). 
Let $d$ be any of the parameters derived from the fit
parameters, $d=d(p_1,\ldots,p_m)$, for instance $\tau_m$ or $\Nnht$. 
For every fit parameter $p^0_k$ ($k=1,\dots,m$) we evaluate the 
values of the derived parameter when we 
increase and decrease the value of the $k$-th fit parameter by its error 
$\sigma(p_k)$, 
\begin{eqnarray}
d^+_k=d(p^0_1,\ldots,p^0_k+\sigma(p_k),\ldots,p^0_m),\quad k=1,\dots, m \nonumber\\
d^-_k=d(p^0_1,\ldots,p^0_k-\sigma(p_k),\ldots,p^0_m),\quad k=1,\dots, m
\end{eqnarray}
Assuming that the errors of the fit parameters are statistically
independent, we can estimate the error  $\sigma(d)$ as
\begin{equation}
\sigma^2(d)= 
\sum_{k=1}^m \left(\frac{d^+_k-d^-_k}{2}\right)^2.
\end{equation}

\section{Comparison of \hfs\ with other routines}
\label{scomp}

\begin{table*}[htb]
\small
\centering
\caption{\label{tcompclass}
Comparison of the results obtained with \hfs\ and CLASS for a sample of
\nh$(1,1)$ spectra.} 
\begin{tabular}{lccccccl}
\tableline\tableline
  &$\Delta V$        &$\vlsr$               &$A(1-\exp\{-\tau_m\})$
  &                  &$A\tau_m$             &                 & \\
Id.\tablenotemark{a}  
  &(\kms)            &(\kms)                &(K)             
  &$1-\exp\{-\tau_m\}$&  (K)                &$\tau_m$              
  &Rout.\tablenotemark{b}\\
\tableline
1 &$1.221\pm0.194$   & $-3.029\pm0.061$     &$\phn6.2\pm0.8$ 
  &  $0.2\pm0.5$     &    $7.1\pm3.5$       &    $0.3\pm0.9$   &\hfs \\
  &$1.240\pm0.091$   & $-3.030\pm0.027$     &\nodata         
  &\nodata           &    $7.1\pm1.0$       &    $0.3\pm0.3$   &C \\[1ex]
2 &$0.712\pm0.098$   & $-4.543\pm0.023$     &   $15.7\pm0.9$ 
  &$0.996\pm0.005$   &$>88$                 & $>5.6$           &\hfs \\
  &$0.750\pm0.007$   & $-4.550\pm0.003$     &\nodata         
  &\nodata           &   $85.9\pm1.6\phn$   &   $5.54\pm0.13$  &C \\[1ex]
3 &$0.569\pm0.066$   &$-20.445\pm0.030\phn$ &$\phn4.5\pm0.4$ 
  &$0.970\pm0.023$   &   $16.2\pm4.8\phn$   &    $3.5\pm1.1$   &\hfs \\
  &$1.030\pm0.001$   &$-20.500\pm0.011\phn$ &\nodata         
  &\nodata           &    $8.0\pm0.4$       &    $1.4\pm0.2$   &C \\[1ex]
4 &$0.199\pm0.023$   &$-20.382\pm0.022\phn$ &$\phn3.4\pm0.3$ 
  &$1.000\pm0.002$   &     $29\pm19$        &    $8.4\pm5.5$   &\hfs \\
  &$1.030\pm0.001$   &$-20.400\pm0.011\phn$ &\nodata         
  &\nodata           &    $9.4\pm0.4$       &   $1.26\pm0.16$  &C \\[1ex]
5 &$3.753\pm0.358$   & $-2.180\pm0.168$     &$\phn6.8\pm0.6$ 
  &  $0.7\pm0.2$     &   $12.3\pm4.0\phn$   &    $1.3\pm0.8$   &\hfs \\
  &$3.410\pm0.039$   & $-2.160\pm0.022$     &\nodata         
  &\nodata           &   $9.87\pm0.10$      &   $0.84\pm0.03$  &C \\
\tableline
\tablenotetext{1}{Spectra analyzed in the comparison. 
1: low optical depth; 
2: very high optical depth; 
3, 4: lines narrower than the channel width (0.6 \kms), moderate and high
optical depths; 
5: very broad lines.}
\tablenotetext{2}{Routines being compared. 
\hfs: present work; 
C: CLASS, using method NH3(1,1)}
\end{tabular}
\end{table*}                                                           

\begin{table*}[htb]
\small
\centering
\caption{\label{tcompros08}
Comparison of the results obtained with \hfs\ and the routines of \citet{Ros08}
for a sample of \nh$(1,1)$ and $(2,2)$ spectra.} 
\begin{tabular}{lccccccccl}
\tableline
     &                &                   &                   &              
     &                &            &\multicolumn{2}{c}{$\Nnht$\tablenotemark{a}}
     &\\
     &$\Delta V\tablenotemark{b}$ &$\vlsr$ &$A\tau_{1m}$\tablenotemark{b} &
     &$\Tex$          &$T_k$       &\multicolumn{2}{c}{($10^{13}$ cm$^{-3}$)}   
     &\\ \cline{8-9}
Id.\tablenotemark{c}   
     &(\kms)               &(\kms)       &(K)     &$\tau_{1m}$\tablenotemark{b}
     &(K)                  &(K)          &($f\ll1$)          &($f=1$)
     &Rout.\tablenotemark{d}\\
\tableline
16   &$0.335\pm0.034$ &$4.596\pm0.013$   &$0.90\pm0.12$      &$1.12\pm0.12$   
     &$3.53\pm0.12$   &$\phn9.9\pm1.9$   &$\phn3.8\pm2.1$    &$17.1\pm9.9$      
     &\hfs\\
     &$0.332\pm0.014$ &$4.614\pm0.007$   &$1.85\pm0.8\phn$   &\nodata         
     &\nodata         &$<13.$            &\nodata            &$>1.27$           
     &R\\[1ex]
31.1 &$0.390\pm0.008$ &$4.618\pm0.005$   &$5.67\pm0.12$      &$1.56\pm0.12$   
     &$6.37\pm0.06$   &$12.6\pm1.7$      &$20.3\pm3.5$       &$35.7\pm6.2$      
     &\hfs\\
     &$0.400\pm0.024$ &$4.6\pm0.2$       &\nodata            &$1.875\pm0.001$ 
     &\nodata         &$11.7\pm0.1$      &\nodata            &$\phn\phd23\pm1.0$
     &R\\[1ex]
31.2 &$0.327\pm0.023$ &$5.984\pm0.016$   &$1.37\pm0.14$      &$1.52\pm0.14$   
     &$3.64\pm0.09$   &$10.6\pm2.3$      &$\phn5.1\pm2.7$    &$\phn21\pm11$ 
     &\hfs\\
     &$0.306\pm0.047$ &$6.01\pm0.05$     &\nodata            &$0.870\pm0.001$ 
     &\nodata         &$10.4\pm0.1$      &\nodata            &$\phn\phd10\pm1.0$
     &R\\[1ex]
47   &$0.443\pm0.007$ &$8.1714\pm0.0020$ &$16.83\pm0.38\phn$ &$4.09\pm0.38$   
     &$6.86\pm0.04$   &$11.49\pm0.30$    &$75.9\pm3.6$       &$126.8\pm5.9\phn$ 
     &\hfs\\
     &$0.444\pm0.001$ &$8.1840\pm0.0007$ &\nodata            &$4.075\pm0.035$ 
     &$7.82\pm0.02$   &$11.69\pm0.04$    &\nodata            &$73.2\pm0.7$      
     &R\\[1ex]
89	 &$0.294\pm0.006$ &$8.1617\pm0.0024$ &$12.00\pm0.35$     &$3.23\pm0.35$ 
     &$6.46\pm0.09$   &$\phn9.9\pm1.0$   &$44.6\pm8.4$       &$\phn78\pm15$ 
     &\hfs\\
	   &$0.294\pm0.002$ &$8.1790\pm0.0010$ &\nodata            &$2.75\pm0.10$ 
     &$7.9\pm0.1$     &$10.5\pm0.1$      &\nodata            &$39\pm1$ 
     &R\\[1ex]
93   &$0.227\pm0.005$ &$6.002\pm0.003$   &$4.93\pm0.23$      &$2.48\pm0.23$   
     &$4.73\pm0.09$   &$\phn9.9\pm0.9$   &$14.1\pm2.4$       &$33.6\pm5.6$      
     &\hfs\\
     &$0.231\pm0.002$ &$6.022\pm0.001$   &\nodata            &$1.65\pm0.10$   
     &$6.1\pm0.2$     &$10.5\pm0.2$      &\nodata            &$\phn\phd14\pm1.0$
     &R\\[1ex]
95   &$0.343\pm0.003$ &$6.044\pm0.002$   &$21.95\pm0.11\phn$ &$6.66\pm0.11$   
     &$6.04\pm0.02$   &$\phn9.70\pm0.01$ &$98.8\pm0.8$       &$181.6\pm1.2\phn$ 
     &\hfs\\
     &$0.342\pm0.002$ &$6.0635\pm0.0009$ &\nodata            &$6.90\pm0.10$   
     &$6.75\pm0.02$   &$10.01\pm0.06$    &\nodata            &$\phd106\pm2.0$
     &R\\
\tableline
\tablenotetext{1}{Beam-averaged \nh\ column densities calculated for 
filling factors $f\ll1$ and $f=1$.}
\tablenotetext{2}{Calculated from the values of $\sigma_V$ and
$\tau_\mathrm{tot}$ given in \citet{Ros08}}
\tablenotetext{3}{Identification of the examples of 
\citet{Ros08}.
16: low optical depth, very weak $(2,2)$; 
31: multicomponent; 
47: high signal-to-noise ratio; 
89: narrow lines;
93: good fit, weak $(2,2)$;
95: high optical depth.}
\tablenotetext{4}{Routines being compared. 
\hfs: present work; 
R: \citet{Ros08}.}
\end{tabular}
\end{table*}

The fits obtained with \hfs\ and \hfsnh\ were compared with those obtained with
other commonly used routines. In Tables \ref{tcompclass} and \ref{tcompros08} we
show some examples of fits performed with \hfs\ and CLASS for the \nh$(1,1)$
line, and the Rosolowsky routine \citep{Ros08} for the simultaneous fit of the
\nh$(1,1)$ and $(2,2)$ lines. The fits shown cover cases of low and high optical
depth, and narrow and wide linewidths. As can be seen in the tables, there is in
general agreement between the \hfs\ results and the other routines. 

Regarding the comparison with CLASS, in the case of linewidths lower than the
channel width (spectra 3 and 4 in Table \ref{tcompclass}), CLASS gives a fixed
value for the linewidth, higher than the channel width and much higher than the
actual linewidth. As a consequence the values of $A\tau_m$ and $\tau_m$ given by
CLASS are scaled down by roughly the same factor, so that CLASS is consistent
with \hfs\ in the values of $A\tau_m\Delta V$ and $\tau_m\Delta V$. In general,
the errors given by CLASS appear to be underestimated.

The comparison with the Rosolowsky routine was performed for the examples of
fits given in \citet{Ros08}. The fitted parameters were taken from Table 3 of
the electronic edition of \citet{Ros08}, and the raw spectral data were
retrieved from  the COMPLETE Web site.\footnote{
\url{https://www.cfa.harvard.edu/COMPLETE/data\_html\_pages/GBT\_NH3.html}}  
For the \hfsnh\ fits, the data were Hanning-smoothed with $\mathtt{HFHW=2}$ and
fitted using $\mathtt{Nksample=400}$ (see below). For the parameters obtained
specifically from the simultaneous fit of the \nh$(1,1)$ and $(2,2)$ lines,
there is a good agreement in the values obtained for $T_k$, and the values of
ammonia column density reported in \citet{Ros08} lie between the two limiting
cases given by \hfsnh, for filling factors $f\ll1$ and $f=1$.

\section{Description of the \hfs\ procedures}
\label{sproc}

While the \hfsnh\ procedures fit the \nh$(1,1)$ and $(2,2)$ transitions
simultaneously,  the general \hfs\ procedures fit a single transition, selected
among those stored in the file 
\texttt{hfs\_transitions.dat}.
This file has to be located in the working directory, or in the directory
pointed at by the environment variable 
\texttt{HFS\_DIR} (see Appendix \ref{ainstall}). 
The first transition in the file is a single line at $\vlsr=0$, useful for
fitting a single Gaussian line. Other transitions in the file include 
\nh$(1,1)$ and $(2,2)$, NH$_2$D$(1_{11},1_{10})$, N$_2$H$^+$(1--0), CN, HCN,
H$^{13}$CN, C$_2$H, C$_2$D, C$^{17}$O,
and more transitions can be added easily to the file. 
The criterion for defining the ``main component'' of a transition is,
in general, all the hyperfines with a velocity offset less than 0.001 \kms, and
for \nh$(1,1)$ and $(2,2)$, less than 0.6 \kms.
Any transition not appearing in the file is assumed to be single.

An optional Hanning smoothing of the spectrum can be performed prior to fitting.
You can select the Hanning filter half-width, \texttt{HFHW}. 
For 
$\mathtt{HFHW=0}$ no smoothing is performed.
$\mathtt{HFHW=1}$ is the standard 3-point Hanning smoothing, and 
the resulting spectrum has half the initial number of channels.
In general, the Hanning smoothing encompasses \texttt{2*HFHW+1} points,  
resulting in a final number of channels \texttt{HFHW+1} times smaller.

The iterative process is controlled by two parameters,
\texttt{Nksample}, the number of thousands of samples of the parameter space,
and
\texttt{Final\_Range}, the ratio of ranges of the last loop and initial search
ranges.
The number of loops is taken as 
$\mathtt{nloop} = \mathtt{Nksample}^{1/2}$,
and the number of seeds and descendants is taken as
$\mathtt{nseed} = \mathtt{ndesc} = (\mathtt{nloop}\times1000)^{1/2}$, 
so that
$\mathtt{nloop}\times\mathtt{nseed}\times\mathtt{ndesc}=
\mathtt{Nksample}\times1000$.
For each loop the ranges will be decreased a factor of 
$f=\mathtt{Final\_Range}^{1/(\mathtt{nloop}-1)}$.

The initial values for the fit are guessed from the intensity, position, and
width of the data peak for the first component, and of the residual (data minus
previous components) for the rest of components (up to a maximum of 9). The
main line optical depth is set arbitrarily to 0.7 ($1-e^{-\tau_m}=0.5$), 
or to an arbitrary low value ($10^{-6}$) for a single Gaussian fit.

Values for the initial search ranges are calculated by the procedures.  
For fitting a single Gaussian line the search range of
$\tau^*_m$ is made 0 to keep it constant.
Additional  constraints for the search ranges of the parameters are
$A^*_m>0$; $0<\tau^*_m<1$; and $\Delta V>\Delta V_\mathrm{min}$, where 
$\Delta V_\mathrm{min}$ is the thermal linewidth for a kinetic temperature 
$\Tk=\Tbg$ of a large molecule
(for the general \hfs\ procedures, 0.025 \kms\ for a mass of 200 $m_\mathrm{H}$),
or of \nh\ (0.086 \kms\ for the \hfsnh\ procedures).

Note that the final value of a fit parameter can be outside the initial range
for the parameter. 
For an initial range $2r$, and a range decreasing factor per loop $f$,
the final value of a parameter can can be up to roughly $r/(1-f)$ apart from its
initial value. 
For instance, for the default values
$\mathtt{Nksample}=200$ (corresponding to $\mathtt{nloop}= 14$), and 
$\mathtt{Final\_Range}= 0.05$, we have $f\simeq0.8$ and $r/(1-f)\simeq5r$.

The different procedures that compose \hfs\ are described in the following.

\subsection{
\texttt{hfs\_fit}, 
\texttt{hfs\_nh3}}

These are interactive graphic procedures for fitting simultaneously multiple velocity
components of an spectral line with hyperfine structure,
or to a pair of \nh$(1,1)$ and $(2,2)$, to spectra read from data files,
and generating files with the synthetic spectra. See an example of a fitting run
of \texttt{hfs\_fit} in Appendix \ref{afitrun} and 
of \texttt{hfs\_nh3} in Appendix \ref{anh3run}.

You can set any number of velocity components, for which the
procedures propose a first guess. Alternatively, you can use the cursor to add
or delete components. The position of a new component is set at the cursor
position, its intensity is the intensity at the cursor position, and the FWHM is
estimated around the cursor position.
The values of the initial search ranges can be changed. Any
of the parameters can be kept constant by setting its range to 0. 

\subsubsection*{Input} 
The spectra to fit are read from ASCII files with a pair of values (velocity,
intensity) per line. Lines beginning with ``!'' or ``\#'' are ignored. The
file(s) can be given as argument(s) to \texttt{hfs\_fit} and \texttt{hfs\_nh3}:
\begin{verbatim}
$ hfs_fit <source>.dat
$ hfs_nh3 <source_11>.dat <source_22>.dat
\end{verbatim}

\subsubsection*{Output} 
\begin{figure*}[thb]
\centering
\includegraphics[width=0.45\textwidth]{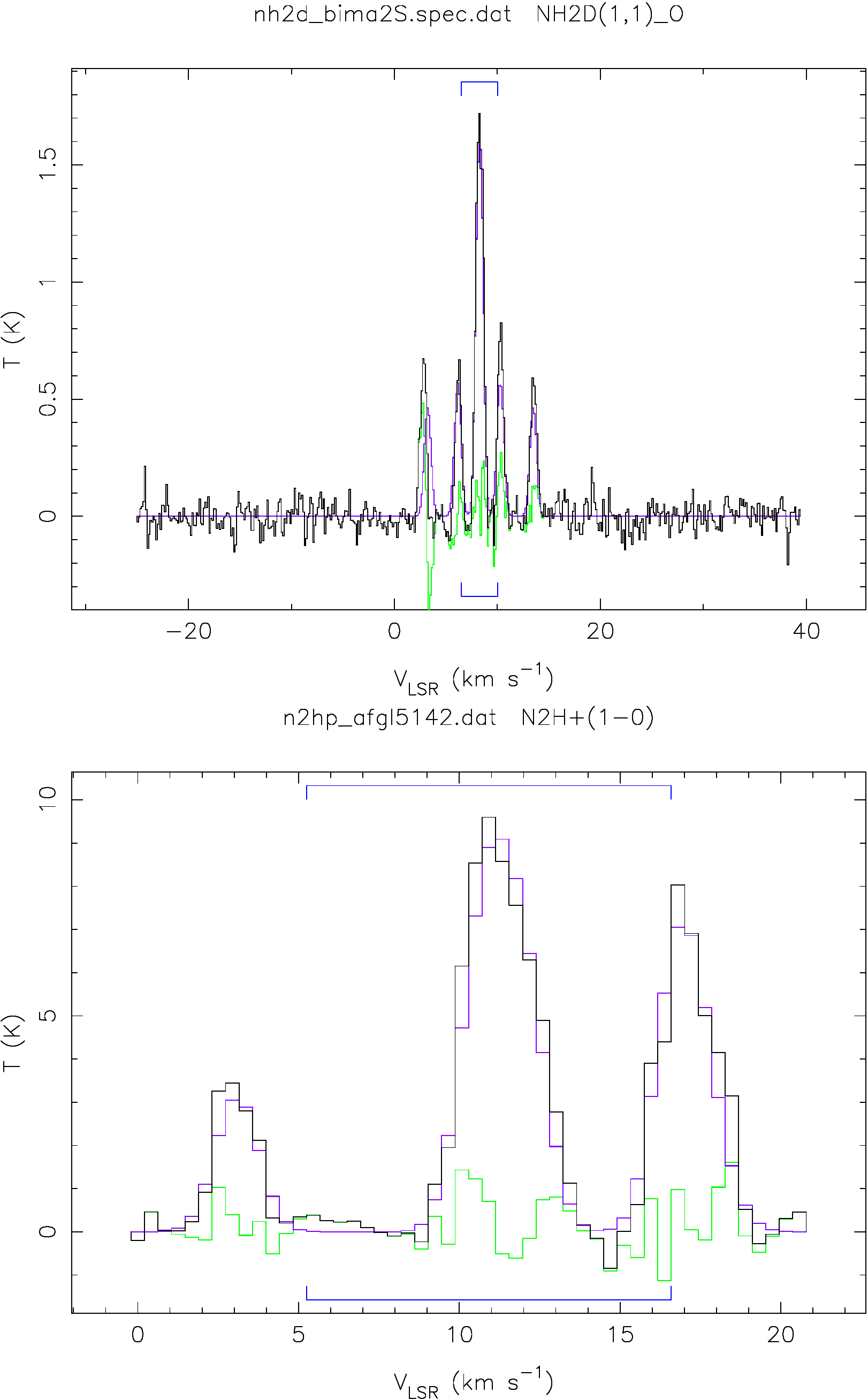}\hfill
\includegraphics[width=0.51\textwidth]{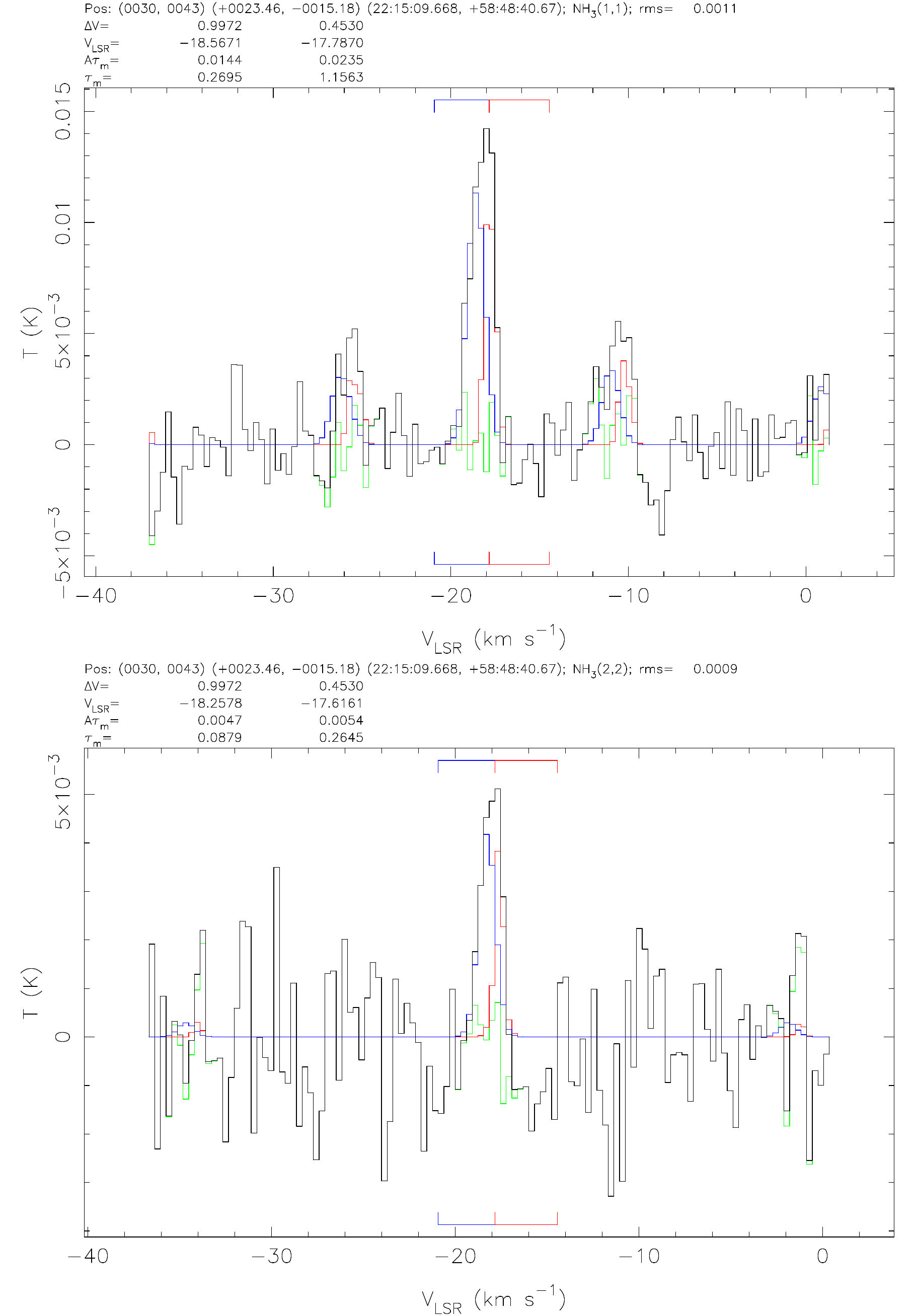}
\caption{\label{ffit}
\emph{Left:} NH$_2$D$(1,1)$ \emph{(top)} and N$_2$H$^+$(1--0) \emph{(bottom)}
fits using \texttt{hfs\_fit}.
The color lines indicate the data \emph{(black)}, fit \emph{(magenta)}, and 
residual \emph{(green)}.
\emph{Right:} \nh$(1,1)$ and $(2,2)$ simultaneous fit of two velocity components
using \texttt{hfs\_nh3}.
The blue and red lines indicate the two velocity components fitted. The search
range for each component is indicated by the blue and red segments at the top
and bottom of each plot.
}
\end{figure*}

\begin{itemize}

\item 
\texttt{hfs\_fit.log} or
\texttt{hfs\_nh3.log},
log file with the details of the fitting session.

\item
\texttt{<source>.synt}, or 
\texttt{<source\_11>.synt} and \texttt{<source\_22>.synt},
ASCII files with the synthesized spectra. 
The files have a header (lines beginning with ``!'') with the values of the
parameters of the fitted spectra for each velocity component. 
The files are readable by GREG of the package
GILDAS.
The file is overwritten for every new fit.
See an example in Appendix \ref{aoutput}.

\item
\texttt{<source>.eps}, plot showing the data, the components fitted, and the
residual. See some examples in Fig.\ \ref{ffit}.  
The file is overwritten for every new fit. 
\end{itemize}

\subsection{
\texttt{hfs\_file}.}

This is a batch procedure to fit the hyperfine structure of the same transition (and a
single velocity component) for a set of data files, and generating a list of
files with the synthetic spectra. The fit procedure is the same as that of
\texttt{hfs\_fit}, but it is not interactive.

\subsubsection*{Input} 
The input is a parameter file that can be given as argument when running the
procedure:
\begin{verbatim}
$ hfs_file <file_list>.par
\end{verbatim}
The contents of the file is as follows:
\begin{enumerate}
\item Transition name, e.g.\ \texttt{"NH3(1,1)"}. 
\item Iteration parameters: \texttt{Nksample}, \texttt{Final\_Range}.
\item and following lines: list of files, one file per line
\end{enumerate}
See an example of \texttt{<file\_list>.par} in Appendix \ref{ainput}.

\subsubsection*{Output} 
\begin{itemize}

\item
\texttt{<file\_list>.log}, 
log file with the
details of the fitting process for all the files in \texttt{<file\_list>.par}.

\item
\texttt{<file\_list>.out}, ASCII file with the values and uncertainties of the
parameters fitted for each file of the list. The first lines beginning with
``!'' are the header and give information about the transition and the column
headers.

\item
\texttt{<file\_list>.ps},
PostScript file with plots of the data, the components
fitted, and the residual for all the files in the list. 

\item
\texttt{<file\_\#>.synt}, an ASCII file for each file in
\texttt{<file\_list>.par}, with the synthesized spectrum. The file has a header
(lines beginning with ``!'') with the values of the parameters of the fitted
spectrum. 

\end{itemize}

\subsection{
\texttt{hfs\_cube\_sp}, 
\texttt{hfs\_nh3\_cube\_sp}}

These are single processor 
batch procedures for fitting simultaneously multiple velocity components of
spectra from 3-axes FITS data cubes. A subimage of the FITS data cube can be
selected, and optional boxcar averaging of pixels and Hanning filtering of the
spectra can be performed. The procedures are similar to \texttt{hfs\_fit} and
\texttt{hfs\_nh3}, but they are not interactive.
The maximum dimension of the image and the maximum number of channels 
of the data cube is arbitrary, i.e.\ the only limitation is the amount of memory 
available by the computer. 

\subsubsection*{Input} 
The input is a parameter file that has to be given as argument when running the
procedures:
\begin{verbatim}
$ hfs_cube_sp <parameter>.par
$ hfs_nh3_cube_sp <parameter>.par
\end{verbatim}
The parameter file for \texttt{hfs\_cube\_sp} has 10 lines
(see an example 
in Appendix \ref{ainput}):
\begin{enumerate}
\item Transition name, e.g.\ \texttt{"C17O(1-0)"}. 
\item FITS data cube file to read. The file must have 3 non-degenerate
axes, in this order: x position, y position, velocity channel.
\item Rms of channels without emission, minimum SNR of spectra to be
fitted. Components with a peak intensity below SNR times rms are not fitted.
\item Number of velocity components to fit, \texttt{Ncomp}, between 1 and 9.
\item Range of channels for each component: $\mathtt{2\times Ncomp}$ values,
with the first and last channel of the velocity  range for each component. 
The channel ranges must be non overlapping.
For the first component, 0 defaults to 1 (first), \texttt{nchan} (last).
\item Hanning filter half-width (channels), 0 for no filtering.
\item Boxcar smoothing radius (pixels), 0 for no smoothing. 
\item Subimage to be fitted: first X pixel, last X pixel, X increment. 
0 defaults to 1 (first), \texttt{ndim1} (last), 1 (increment).
\item Subimage to be fitted: first Y pixel, last Y pixel, Y increment. 
0 defaults to 1 (first), \texttt{ndim2} (last), 1 (increment).
\item Iteration parameters: \texttt{Nksample}, \texttt{Final\_Range}.
\end{enumerate}

The parameter file for
\texttt{hfs\_nh3\_cube\_sp} has 11 lines
(see an example 
in Appendix \ref{ainput}):
\begin{enumerate}
\item $(1,1)$ FITS data cube file to read. The file must have 3 non-degenerate
axes, in this order: x position, y position, velocity channel.
\item $(1,1)$ rms of channels without emission, minimum SNR of spectra to be
fitted. Components with a peak intensity below SNR times rms are not fitted.
\item Number of velocity components to fit, \texttt{Ncomp}, between 1 and 9.
\item $(1,1)$ range of channels for each component: $\mathtt{2\times Ncomp}$
values, with the first and last channel of the velocity  range for each
component. The channel ranges must be non overlapping. For the first component,
0 defaults to 1 (first), \texttt{nchan} (last).
\item $(2,2)$ FITS data cube file to read. The file must have the same geometry
as that of the $(1,1)$ FITS file.
\item $(2,2)$ rms of channels without emission, minimum SNR of spectra to be
fitted. Components with a peak intensity below SNR times rms are not fitted.
\end{enumerate}
The 5 following lines are the same as for the case of \texttt{hfs\_cube\_sp}.

\subsubsection*{Output}
\begin{itemize}

\item
\texttt{log/<parfile>.log}, 
log file in folder \texttt{log} with the
details of the fitting process for all the pixels of the subimage.

\item
\texttt{<parfile>\_comp\#.out},  
an ASCII file for each velocity component with the values of the parameters
fitted and the line and physical parameters for each pixel of the subimage, and
their uncertainty. 
The first lines beginning with "!" are the header and give information about the
parameter file, FITS files, velocity component number, velocity range of the
component, Hanning filtering applied, smoothing boxcar radius, 
and column headers. 
The files are used by \texttt{hfs\_view}, and are readable by GREG.
See an example in Appendix \ref{aoutput}.

\item
\texttt{ps/<parfile>\_<xoffset>.ps},
PostScript files in folder \texttt{ps}, with plots of the data, the components
fitted, and the residual for all pixels of the subimage with a given
\texttt{<xoffset>} in arcsec. 

\begin{figure*}[thb]
\centering
\plotone{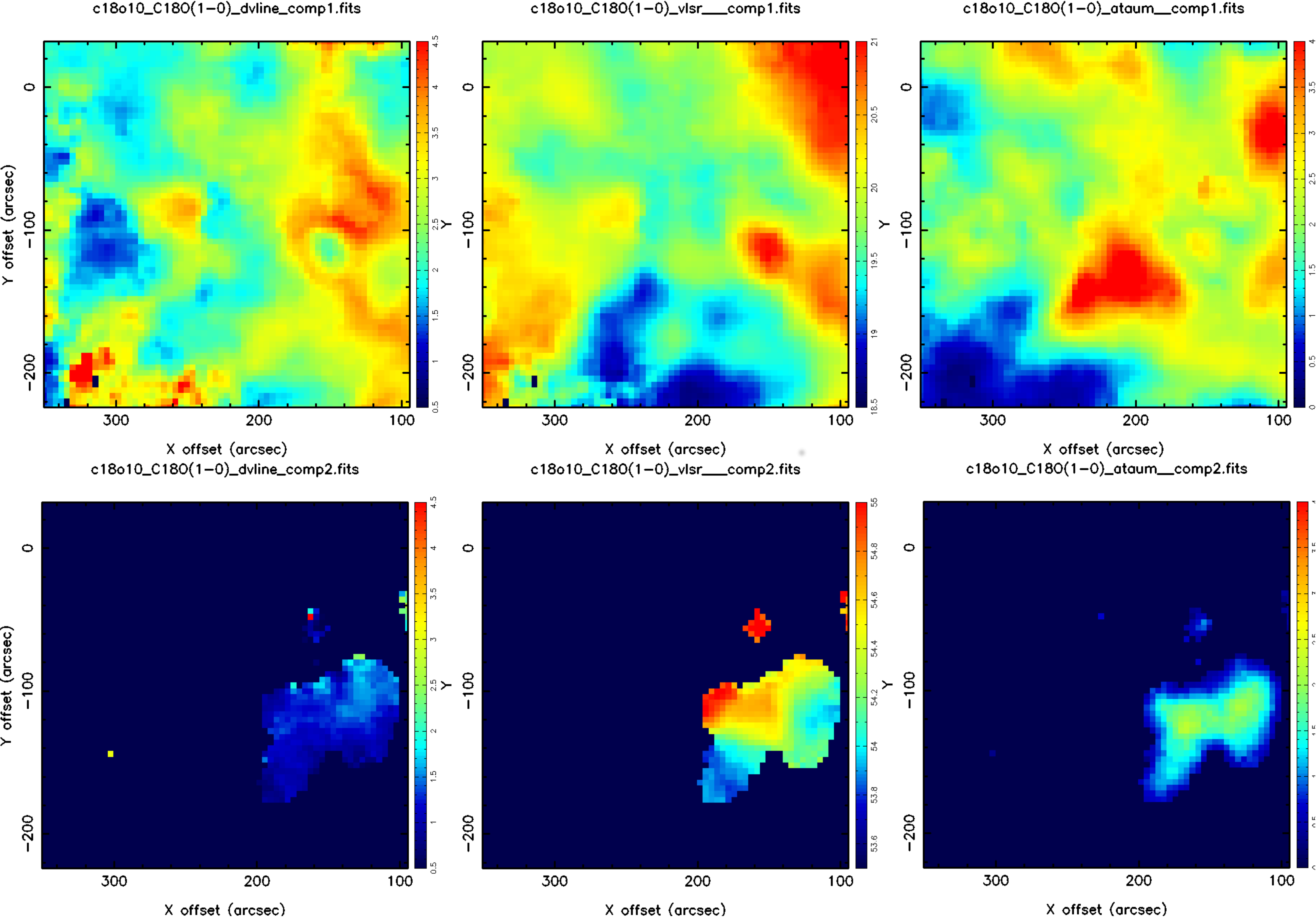}
\caption{\label{fcube}
Example of maps of the parameters fitted with \texttt{hfs\_cube} of two velocity
components to  the C$^{18}$O (1--0) transition for a $64\times64$ data cube. 
\emph{Top:} component 1, 
\emph{bottom:} component 2, 
\emph{left:} $\Delta V$,
\emph{middle:} $\vlsr$,
\emph{right:} $A\tau_m$.
}
\end{figure*}

\item
\texttt{maps/<parfile>\_<parameter>\_comp\#.fits},
FITS files in folder \texttt{maps}, with maps, for each velocity component, of
the parameters fitted and the line and physical parameters,  and their
uncertainty. 
Each FITS file has two planes, the first plane with the values of the parameter,
and the second plane with the uncertainties.
See in Fig.\ \ref{fcube} an example of the the maps obtained from a
2-velocity-components fit of the C$^{18}$O line for a FITS data cube.
\end{itemize}

\subsection{
\texttt{hfs\_cube\_mp},
\texttt{hfs\_nh3\_cube\_mp}}

These are multiprocessor procedures that use Open MPI (see Appendix \ref{ainstall})
and run in parallel using a number of processors available in the machine, or
in more than one host. The multiprocessor procedures
are naked versions of the single-processor versions, 
\texttt{hfs\_cube\_sp} and \texttt{hfs\_nh3\_cube\_sp}, 
without any graphic output. The instructions for running these procedures can be
found in Appendix \ref{amp}.

\subsection{
\texttt{hfs\_view},
\texttt{hfs\_nh3\_view}}

These are interactive graphic procedures for displaying spectra from FITS data cubes and
the corresponding synthetic spectra fitted with  
\texttt{hfs\_cube} or
\texttt{hfs\_nh3\_cube}. 
A plot of the integrated intensity is shown, and you can select with the mouse
the position for which the the spectrum (data and synthetic) is shown.
You can select to show the integrated intensity for all channels, or for the
channel ranges of each velocity component.
The data and synthetic spectra at any pixel can be extracted in ASCII files.

\subsubsection*{Input} 
The input is the same parameter file used as input to 
\texttt{hfs\_cube} or
\texttt{hfs\_nh3\_cube}. 
From the information in this file, 
\texttt{hfs\_view}
reads the corresponding FITS data cubes and the 
\texttt{<parfile>\_comp\#.out} 
file for each velocity component (created by 
\texttt{hfs\_cube} or \texttt{hfs\_nh3\_cube}),
with the parameters of the fitted spectra.

\subsubsection*{Output} 
For any pixel you can extract the data and synthetic spectra:
\begin{itemize}

\item 
\texttt{<parfile>\_<xoffset>\_<yoffset>.spec}, or \\
\texttt{<parfile>\_11\_<xoffset>\_<yoffset>.spec} and  
\texttt{<parfile>\_22\_<xoffset>\_<yoffset>.spec},
data spectra for the pixel selected, with the Hanning filtering and boxcar
smoothing given in the \texttt{<parfile>.par} file.
See an example in Appendix \ref{aoutput}.

\item
\texttt{<parfile>\_<xoffset>\_<yoffset>.synt}, or \\
\texttt{<parfile>\_11\_<xoffset>\_<yoffset>.synt} and 
\texttt{<parfile>\_22\_<xoffset>\_<yoffset>.synt},
synthetic spectra fitted for the pixel selected.

\end{itemize}

\subsection{
\texttt{hfs\_blanking}} 

This is an auxiliary procedure to flag the output FITS files of 
\texttt{hfs\_cube} and 
\texttt{hfs\_nh3\_cube},
according to the parameter values, errors, or SNR.
The FITS files 
have two planes, the first one with the values of the parameter at each pixel,
and the second one with with the error of the value.
The procedure reads the values and errors in the FITS files, 
and allows you to blank the pixels
that fulfill a criterion on parameter error, value, or value/error, above or
below a cutoff value.
 
\subsubsection*{Input} 
\texttt{<parfile>\_<parameter>\_comp\#.fits}, 
any FITS file created by
\texttt{hfs\_cube} or 
\texttt{hfs\_nh3\_cube},
for each velocity component and parameter. 

\subsubsection*{Output} 
\texttt{<parfile>\_<parameter>\_comp\#\_blank.fits},
output 2-axes FITS file,
with the values of the parameter for
non-flagged pixels, and NaN for flagged-out pixels.

\subsection{
\texttt{hfs\_synt}} 

Auxiliary procedure to create a synthetic spectrum, with the option
of adding Gaussian noise. You can select the transition, number of channels,
spectral resolution, line parameters, and noise level.

\subsubsection*{Output} 
\begin{itemize}

\item
\texttt{hfs\_synt.synt}, ASCII file with the synthetic spectrum generated.
The file has a header (lines beginning with ``!'') with the values of the
parameters used for generating the synthetic spectrum. 

\item
\texttt{hfs\_synt.eps}, plot of the synthetic spectrum.

\end{itemize}

\acknowledgements

The author thanks 
Pau Estalella for suggesting the use of pseudo-random sequences,
Ferran Sala for helpful discussions on the projections of a quadric,
and
Salvador Curiel for helping with the implementation of the multiprocessor
procedures.
Thanks also to 
\'Alvaro S\'anchez-Monge for reading the manuscript and, together with
Aina Palau, Gemma Busquet and Carmen Ju\'arez, for
testing \hfs, finding bugs, and suggesting improvements.
This work has been partially supported by the Spanish MINECO grant
AYA2014-57369-C3 (cofunded with FEDER funds) and
MDM-2014-0369 of ICCUB (Unidad de Excelencia `Mar\'{\i}a de Maeztu').

\bibliographystyle{apj}

\begin{thebibliography}{}

\bibitem[Anglada et al.\ (1995)]{Ang95}
Anglada, G., Estalella, R., Mauersberger, R., Torrelles, J. M., Rodr\'{\i}guez,
L. F., Cant\'o, J., Ho, P. T. P., D'Alessio, P. 1995,  ApJ, 443, 682 

\bibitem[Avni (1976)]{Avn76}
Avni, Y. 1976, ApJ, 210, 642

\bibitem[Cant\'o, Curiel, Mart\'{\i}nez-G\'omez (2009)]{Can09}
Cant\'o, J., Curiel, S., Mart\'{\i}nez-G\'omez, E. 2009, A\&A, 501, 1259

\bibitem[Danby et al.\ (1988)]{Dan88}
Danby, G., Flower, D. R., Valiron, P., Schilke, P., \& Walmsley, C. M. 1988
MNRAS, 235, 229

\bibitem[Estalella (2016)]{Est16}
Estalella, R. 2016, HfS: Hyperfine Structure fitting tool, 
Astrophysics Source Code Library, record ascl:1607.011

\bibitem[Estalella \& Anglada (1997)]{Est97}
Estalella, R., Anglada, G. 1997,  ``Introducci\'on a la F\'{\i}sica del Medio
Interestelar'',  Col$\cdot$lecci\'o Textos Docents, n.\ 50, 2nd edition: 
Edicions
de la Universitat de Barcelona, Spain

\bibitem[Estalella et al.\ (2012)]{Est12}
Estalella, R., L\'opez, R, Anglada, G., G\'omez, G., Riera, A.,
Carrasco-Gonz\'alez, C. 2012,  AJ, 144, 61

\bibitem[Faure et al.\ (2013)]{Fau13}
Faure, A., Hily-Blant, P., Le Gal, R., Rist, C., Pineau des For\^ets, G.
2013, ApJ, 770, L2
	
\bibitem[Goddi, Zhang, \& Moscadelli (2015)]{God15}
Goddi, C., Zhang, Q., Moscadelli, L. 2015, A\&A, 573, A108

\bibitem[Halton (1964)]{Hal64}
Halton, J. H. 1964, Commun.\ ACM, 7, 701

\bibitem[Ho \& Townes (1983)]{Ho83}
Ho, P. T. P., Townes, C. H. 1983, ARAA, 21, 239

\bibitem[Kukolich (1967)]{Kuk67}
Kukolich, S. G. 1967, Phys.\ Rev., 156, 83


\bibitem[Mangum \& Shirley (2015)]{Man15}
Mangum, J. G., Shirley, Y. L. 2015, PASP, 127, 266

\bibitem[Maret et al.\ (2009)]{Mar09}
Maret, S., Faure, A., Scifoni, E., \& Wiesenfeld, L. 2009, MNRAS, 399, 425

\bibitem[McConnell (2011)]{McC11}
McConnell, A. J. 2011, ``Applications of Tensor Analysis'',  Dover Publications

\bibitem[Osorio et al.\ (2009)]{Oso09}
Osorio, M., Anglada, G., Lizano, S., D'Alessio, P. 2009, ApJ, 694, 29

\bibitem[Palau et al.\ (2014)]{Pal14}
Palau, A., Estalella, R., Girart, J. M., Fuente, A., Fontani, F., Commer\c{c}on,
B., Busquet, G., Bontemps, S., S\'anchez-Monge, \'A., Zapata, L. A, Zhang, Q.,
Hennebelle, P., di Francesco, J. 2014,  ApJ, 785, 42

\bibitem[Poynter \& Kakar (1975)]{Poy75}
Poynter, R. L., Kakar, R. K. 1975, ApJ, 29, 87	

\bibitem[Rosolowsky et al.\ (2008)]{Ros08}
Rosolowsky, E. W., Pineda, J. E., Foster, J. B., Borkin, M. A., Kauffmann, J., 
Caselli, P., Myers, P. C., Goodman, A. A. 2008, ApJSS, 175, 509

\bibitem[S\'anchez-Monge et al.\ (2013)]{San13}
S\'anchez-Monge, \'A., Palau, A., Fontani, F., Busquet, G., Ju\'arez, C.,
Estalella, R., Tan, J. C., Sep\'ulveda, I., Ho, P. T. P., Zhang, Q., Kurtz S.
2013, MNRAS, 432, 3288

\bibitem[Sep\'ulveda et al.\ (2011)]{Sep11}
Sep\'ulveda, I., Anglada, G., Estalella, R., L\'opez, R., Girart, J.M., Yang, J.
2011, A\&A, 527, A41 

\bibitem[Svoboda et al.\ (2016)]{Svo16}
Svoboda, B. E., Shirley, Y. L., Battersby, C., Rosolowsky, E. W.,  Ginsburg, A.
G., Ellsworth-Bowers, T. P., Pestalozzi, M. R., Dunham, M. K.,  Evans, N. J.,
II, Bally, J., Glenn, J. 2016, ApJ, 822, 59

\bibitem[Sobol (1967)]{Sob67}
Sobol, I. 1967, USSR Computational Mathematics and Mathematical Physics, 7, 86

\bibitem[Swift et al.\ (2005)]{Swi05}
Swift, J. J., Welch, W. J., \& Di Francesco, J. 2005
ApJ, 620, 823

\bibitem[Walmsley \& Ungerechts (1983)]{Wal83}
Walmsley, C. M. \& Ungerechts, H., 1983. A\&A, 122, 164

\bibitem[Wall \& Jenkins (2003)]{Wal03}
Wall, J. V. \& Jenkins, C. R. 2003, ``Practical Statistics for Astronomers'',
Cambridge University Press

\end{thebibliography}

\appendix

\section{A. \hfs\ requisites and installation}
\label{ainstall}

The \hfs\ procedures run on a Linux or Mac OS X system with a Fortran 90
compiler (for instance \texttt{gfortran}), and use the 
PGplot Graphics Subroutine Library 
compiled with \texttt{gfortran} (see Appendix \ref{apgplot}), 
for the graphic output, and  Open MPI 
for the multiprocessor procedures \texttt{hfs\_cube\_mp} and 
\texttt{hfs\_nh3\_cube\_mp}.

\hfs\ can be freely downloaded as a file \texttt{hfs.tgz} from the
Astrophysics Source Code Library, 
record ascl:1607.011\footnote{\url{http://ascl.net/1607.011}}
Once you have downloaded \texttt{hfs.tgz},
untar the file in your installation directory (as root or using
\texttt{sudo}), for instance \texttt{/usr/local/hfs},
\begin{verbatim}
$ mkdir /usr/local/hfs
$ mv hfs.tgz /usr/local/hfs
$ cd /usr/local/hfs
$ tar -xzvf hfs.tgz
\end{verbatim}
If necessary, edit the first lines of the shell script \texttt{hfs\_compile} to
change the lines
\begin{verbatim}
compiler="gfortran"
libraries="-lpgplot -lX11"
\end{verbatim}
You may need  to give the location of the PGplot library with the option
\begin{verbatim}
libraries="-lpgplot -lX11 -L/usr/local/lib/pgplot"
\end{verbatim}
Mac OS X users may need to indicate the location of the X11 library too
\begin{verbatim}
libraries="-lpgplot -lX11 -L/usr/local/lib/pgplot -L/usr/X11/lib"
\end{verbatim}
Run the compile script
\begin{verbatim}
$ ./hfs_compile
\end{verbatim}
The shell script \texttt{hfs\_links} create symbolic links in 
\texttt{\$exe\_dir} pointing at the HfS procedures. 
If necessary, change the line
\begin{verbatim}
exe_dir="/usr/local/bin"
\end{verbatim}
and run the links script
\begin{verbatim}
$ ./hfs_links
\end{verbatim}
Define the environment variable \texttt{HFS\_DIR}, pointing at your installation
directory. It is used by \hfs\ to find the files
\texttt{hfs\_transitions.dat} and
\texttt{hfs\_fit.help}. This can be done by adding to your \texttt{.bashrc} or
\texttt{.bash\_profile} the lines
\begin{verbatim}
# hfs
export HFS_DIR=/usr/local/hfs
\end{verbatim}

\section{B. Installation of PGplot with gfortran}
\label{apgplot}

Once you have downloaded the PGplot distribution file,
follow the normal installation procedure (as root or using
\texttt{sudo}):
\begin{verbatim}
$ mv pgplot5.2.tar.gz /usr/local/src
$ cd /usr/local/src
$ tar xzvf pgplot5.2.tar.gz
$ mkdir /usr/local/pgplot
$ cd /usr/local/pgplot
$ cp /usr/local/src/pgplot/drivers.list .
\end{verbatim}
Edit the file \texttt{drivers.list} and uncomment (select) the drivers\\
\texttt{/NULL} (null device),\\ 
\texttt{/PS, /VPS, /CPS, /VCPS}  (PostScript drivers), and\\ 
\texttt{/XWINDOW, /XSERVE} (X window drivers).\\ 
Run \texttt{makemake} to prepare the makefile for a Linux system with
\texttt{g77\_gcc} compiler,
\begin{verbatim}
$ /usr/local/src/pgplot/makemake /usr/local/src/pgplot linux g77_gcc
\end{verbatim}
Edit the file \texttt{makefile} and replace lines 25 and 26:
\begin{verbatim}
FCOMPL=g77
FFLAGC=-u -Wall -fPIC -O
\end{verbatim}
by the following lines:
\begin{verbatim}
FCOMPL=gfortran
FFLAGC=-ffixed-form -ffixed-line-length-none -u -Wall -fPIC -O
\end{verbatim}
Continue the normal installation procedure,
\begin{verbatim}
$ make
$ make clean
\end{verbatim}
and, assuming that \texttt{LD\_LIBRARY\_PATH} points at \texttt{/usr/local/lib},
\begin{verbatim}
$ cd /usr/local/lib
$ ln -s /usr/local/pgplot/libpgplot.so .
\end{verbatim}
Define the environment variable \texttt{PGPLOT\_DIR}, pointing at your installation
directory. This can be done by adding to your \texttt{.bashrc} or
\texttt{.bash\_profile} file the lines
\begin{verbatim}
# PGplot
export PGPLOT_DIR='/usr/local/pgplot'
\end{verbatim}
Additionally, you can customize PGplot by adding the following lines to your
\texttt{.bashrc} or \texttt{.bash\_profile}:
\begin{verbatim}
# Default xwindow device
export PGPLOT_DEV='/xwin'
# Default white background
export PGPLOT_BACKGROUND='white'
# Default black foreground
export PGPLOT_FOREGROUND='black'
# Marking text written in the ps file so it can be be edited
export PGPLOT_PS_VERBOSE_TEXT='yes'
# Starting pgxwin_server with 256 colors, server window not visible
/usr/local/pgplot/pgxwin_server -win_maxColors 256 -server_visible False
\end{verbatim}

\section{C. Running the multiprocessor procedures 
\texttt{hfs\_cube\_mp} and \texttt{hfs\_nh3\_cube\_mp}}
\label{amp}

You need to have Open MPI installed to be able to run 
\texttt{hfs\_cube\_mp} or \texttt{hfs\_nh3\_cube\_mp}.
To run the procedures, type
\begin{verbatim}
$ mpirun -np <N> hfs_cube_mp <parfile>.par
$ mpirun -np <N> hfs_nh3_cube_mp <parfile>.par
\end{verbatim}
where \texttt{<N>} is the number of processors to use.
To know the number (and characteristics) of processors in a Linux system, 
you can type
\begin{verbatim}
$ cat /proc/cpuinfo
\end{verbatim}
For running in more than one host, for example, in
\texttt{localhost} and another host
\texttt{<otherhost>}, type
\begin{verbatim}
$ mpirun -np <N> -host localhost,<otherhost> hfs_cube_mp <parfile>.par
$ mpirun -np <N> -host localhost,<otherhost> hfs_nh3_cube_mp <parfile>.par
\end{verbatim}
Here \texttt{<N>} is the total number of processors to use, distributed
among the hosts listed after \texttt{-host}. 
If you want to know which are the processes run in each host, add the option
\texttt{-display-map}.

Requisites for running in a remote host \texttt{<otherhost>}:
\begin{itemize}

\item 
\texttt{ssh} access to \texttt{<otherhost>}, 
without having to enter the password, i.e.\ with your local
\texttt{id\_rsa\_pub} added to \texttt{<otherhost>:.ssh/authorized\_keys2}. 

\item 
Open MPI installed in \texttt{<otherhost>}. 
Your \texttt{PATH} and \texttt{LD\_LIBRARY\_PATH} in
\texttt{<otherhost>} have to point at the openmpi 
\texttt{bin} and \texttt{lib} folders in the
installation directory. For example, if openmpi is installed in
\texttt{/usr/local/openmpi}, you can include in the file 
\texttt{<otherhost>:.bashrc} 
the lines
\begin{verbatim}
export PATH=$PATH:/usr/local/openmpi/bin
export LD_LIBRARY_PATH=$LD_LIBRARY_PATH:/usr/local/openmpi/lib
\end{verbatim}

\item 
The same data file structure in \texttt{<otherhost>} and in \texttt{localhost},
i.e.\ the same directory from where you run \texttt{hfs\_cube\_mp} or
\texttt{hfs\_nh3\_cube\_mp}, with the same  parameter file
\texttt{<parfile>.par} and data files. The log files of the different
processors, \texttt{log/<parfile>\_\#\#.log}, will be written in the host where
each processor runs.

\end{itemize}

\section{D. Example of a \texttt{hfs\_fit} run}
\label{afitrun}

\small
\begin{verbatim}
$ hfs_fit nh311_thin.dat
_____________________________________________
HfS_fit. HyperFine Spectra multicomponent fit
Robert Estalella, 2015/06
_____________________________________________
******  Use a terminal with at least 94 columns  ******
Date: 2016/03/22 Time: 12:45:04

Data file: nh311_thin.dat                                                  
N. of data points read:         87
Reference channel (V=0):   59.6269
Channel width (km s^-1):    0.3089
Off-line rms:               0.2104
Transition:             NH3(1,1)                        
tau_tot/tau_m:              2.0000

Fit rms:       0.3933
Present values and search ranges (ncomp= 1)
Param:        Delta_V       V_lsr A(1-e^-t_m)  1-e^-tau_m
Comp:   1 _______________________________________________
Value:         1.2103     -2.9734      6.1716      0.5000
Range:         1.2103      1.2103      0.3841      0.5000

HfS fit menu_______________________________________
0. Quit
1. Help
2. Read data file:    nh311_thin.dat                                                  
3. Select transition: NH3(1,1)                        
4. Hanning smoothing
5. Select plot (1:data +2:comp +4:res +8:synt):   7
6. Change Nksample, Final_Range:       200    0.050
7. Change Ncomp= 1 and make initial guess
8. Change initial values or ranges
9. Fit 1 component(s) and estimate errors
Choose option (0-9): 9

Iteration parameters read
Nksample:                200
Final_Range:           0.050
Iteration parameters used
Nseed:                   118
Ndesc:                   118
Nloop:                    14
Range_Fact:            0.794
Fitting ncomp= 1 component(s)
Loop Comp     Delta_V       V_lsr A(1-e^-t_m)  1-e^-tau_m         rms
_________ ___________________________________________________________
   0    1      1.2103     -2.9734      6.1716      0.5000      0.3933
   1    1      2.1267     -3.9476      6.3392      0.9930      0.3933
   2    1      1.4883     -2.9911      5.8081      0.3219      0.3784
   3    1      1.1926     -3.1046      6.1852      0.2803      0.3766
   4    1      1.2946     -2.9713      6.0164      0.1679      0.3762
   5    1      1.2304     -3.0472      5.9933      0.1664      0.3762
   6    1      1.1915     -2.9995      6.3971      0.2483      0.3762
   7    1      1.2597     -3.0325      5.9275      0.2578      0.3762
   8    1      1.1783     -3.0519      6.2819      0.3125      0.3762
   9    1      1.1862     -3.0152      6.2391      0.3103      0.3761
  10    1      1.2051     -3.0217      6.3489      0.1883      0.3761
  11    1      1.2628     -3.0253      6.1733      0.1857      0.3760
  12    1      1.2043     -3.0294      6.2455      0.2270      0.3760
  13    1      1.2196     -3.0238      6.1633      0.2602      0.3760
  14    1      1.2284     -3.0299      6.1945      0.2700      0.3760

Error estimation in progress
Fit rms:          0.3760
N. fitted par:         4
Target rms:       0.3865
Single parameters........
Pairs of parameters.........
Par   Intersect  Projection
  1      0.1105      0.1938
  2      0.0585      0.0608
  3      0.4400      0.7270
  4      0.2665      0.5207
Derived parameters
Done

Fit rms:       0.3760
Best fit and errors______________________________________ Derived parameters_________________
Param:        Delta_V       V_lsr A(1-e^-t_m)  1-e^-tau_m      Atau_m       tau_m           A
Comp:   1 _______________________________________________ ___________________________________
Value:         1.2188     -3.0303      6.2258      0.2479      7.1547      0.2849  2.5114E+01
Error:         0.1938      0.0608      0.7270      0.5207      3.4903      0.8576  4.4023E+06
\end{verbatim}
\normalsize

\section{E. Example of a \texttt{hfs\_nh3} run}
\label{anh3run}

\small
\begin{verbatim}
$ hfs_nh3 nh3_11.dat nh3_22.dat
___________________________________________________
NH3 (1,1) and (2,2) multicomponent fit and analysis
Robert Estalella. 2015/12
___________________________________________________
******  Use a terminal with at least 118 columns  ******
Date: 2016/03/22 Time: 12:04:18

Data file 1: nh3_11.dat
N. of data points read:        413
Reference channel (V=0):  208.0598
Channel width (km s^-1):    0.2059

Data file 2: nh3_22.dat
N. of data points read:        413
Reference channel (V=0):  207.9984
Channel width (km s^-1):    0.2057

Fit rms 1,2:   0.0582      0.0157
Present values and search ranges (Ncomp= 1)
(J,K):    (1,1)&(2,2)       (1,1)       (1,1)       (1,1)       (2,2)       (2,2)
Param:        Delta_V       V_lsr A(1-e^-t_m)  1-e^-tau_m       V_lsr A(1-e^-t_m)
Comp:   1 _______________________________________________________________________
Value:         0.9006      0.3998      1.5594      0.5000      0.4120      0.2892
Range:         0.9006      1.0000      0.0580      0.5000      1.0000      0.0157

HfS NH3(1,1)&(2,2) fit menu________________________
0. Quit
1. Help
2. Read data files: nh3_11.dat, nh3_22.dat
3. Physical parameter estimation
4. Hanning smoothing
5. Select plot (1:data +2:comp +4:res +8:synt):  15
6. Change Nksample, Final_Range:       200    0.050
7. Change Ncomp= 1 and make initial guess
8. Change initial values or ranges
9. Fit 1 component(s) and estimate errors
Choose option (0-9): 9

Iteration parameters read
Nksample:                200
Final_Range:           0.050
Iteration parameters used
Nseed:                   118
Ndesc:                   118
Nloop:                    14
Range_Fact:            0.794

Fitting ncomp= 1 component(s)
Loop Comp     Delta_V      V_lsr1 A(1-e^-t1m)  1-e^-tau1m      V_lsr2 A(1-e^-t2m)     rms_tot
_____________________________________________________________________________________________
   0    1      0.9006      0.3998      1.5594      0.5000      0.4120      0.2892      0.0603
   1    1      1.0775     -0.4231      1.5179      0.9924      1.3918      0.3027      0.0543
   2    1      0.8312      0.4102      1.5944      0.6611      1.3326      0.2791      0.0351
   3    1      0.8878      0.3940      1.6405      0.8201      0.5550      0.2785      0.0310
   4    1      0.7420      0.4763      1.6033      0.8780      0.0978      0.3075      0.0310
   5    1      0.5446      0.3985      1.5672      0.9798      1.9346      0.3106      0.0305
   6    1      0.6609      0.4237      1.5769      0.9845      0.3772      0.2785      0.0305
   7    1      0.5775      0.3836      1.5560      0.9914      0.5027      0.2989      0.0294
   8    1      0.5718      0.4494      1.7124      0.9730      0.2922      0.2766      0.0287
   9    1      0.6237      0.4140      1.5783      0.9651      0.4681      0.3009      0.0285
  10    1      0.5559      0.4449      1.4869      0.9923      0.2012      0.3003      0.0285
  11    1      0.5236      0.4082      1.5732      0.9881      0.4519      0.2765      0.0285
  12    1      0.5844      0.4423      1.5001      0.9866      0.4943      0.3013      0.0283
  13    1      0.5683      0.4253      1.5108      0.9860      0.2885      0.2976      0.0282
  14    1      0.5828      0.4155      1.4899      0.9862      0.4128      0.3338      0.0282

Error estimation in progress
Fit rms:          0.0282
N. fitted par:         6
Target rms:       0.0283
Single parameters............
Pairs of parameters....................
Par   Intersect  Projection
  1      0.0144      0.0361
  2      0.0094      0.0099
  3      0.0318      0.0000
  4      0.0017      0.0000
  5      0.0754      0.0754
  6      0.0696      0.0725
Derived parameters
Done

Fit rms 1,2:   0.0232      0.0164
Best fit and errors______________________________________________________________
(J,K):    (1,1)&(2,2)       (1,1)       (1,1)       (1,1)       (2,2)       (2,2)
Param:        Delta_V       V_lsr A(1-e^-t_m)  1-e^-tau_m       V_lsr A(1-e^-t_m)
Comp:   1 ___________ ___________________________________ _______________________
Value:         0.5560      0.4218      1.4801      0.9910      0.4158      0.3228
Error:         0.0361      0.0099      0.0318      0.0017      0.0754      0.0725

Derived parameters___________________________________________________
(J,K):          (1,1)       (1,1)       (2,2)       (2,2) (1,1)&(2,2)
Param:         Atau_m       tau_m      Atau_m       tau_m           A
Comp:   1 _______________________ _______________________ ___________
Value:         7.0371      4.7118      0.3638      0.2436  1.4935E+00
Error:         0.3068      0.1865      0.0927      0.0623  3.2230E-02

Physical parameters__________________________________________________________________________________________________
Param:      (f=1)T_ex       T_rot         T_k  (f<1)N(11)  (f=1)N(11)  (f<1)N(22)  (f=1)N(22) (f<1)N(NH3) (f=1)N(NH3)
                  (K)         (K)         (K)     (cm^-2)     (cm^-2)     (cm^-2)     (cm^-2)     (cm^-2)     (cm^-2)
Comp:   1 ___________ _______________________ _______________________ _______________________ _______________________
Value:         4.2276      9.6960      9.9724  1.0236E+14  2.9150E+14  2.4889E+12  7.0877E+12  4.5732E+14  1.3023E+15
Error:         0.0324      0.5968      0.6588  8.0026E+12  2.2179E+13  6.5442E+11  1.8673E+12  6.5753E+13  1.8463E+14
\end{verbatim}
\normalsize

\section{F. Examples of \hfs\ input files}
\label{ainput}

\subsection{\texttt{<file\_list>.par}}
\small
\begin{verbatim}
"NH3(1,1)"                    ! Transition           
400 0.05                      ! Nksample, Final_Range
00117+6412.nh311.clump1.spt   ! 1st data file      
AFGL5142.nh311.clump1.dat     ! 2nd data file
CepA.nh311.clump1.spt         ! ...
ON1.nh311.clump1.spt          ! Last data file  
\end{verbatim}
\normalsize

\subsection{
Parameter file for \texttt{hfs\_cube\_sp} and \texttt{hfs\_cube\_mp}}
\small
\begin{verbatim}
"N2H+(1-0)"                   ! Transition
"n2hp-merged-lmv-clean.fits"  ! Input data cube file
0.06 8.0                      ! Rms, minimum SNR to analyze the spectrum
1                             ! Number of components
104 108                       ! Range of channels to search for peak
0                             ! Hanning filter half-width (chan): 0=no; >0=yes
0                             ! Boxcar smoothing radius (pixels): 0=no; >0=yes
20 236 4                      ! Xpix_ini, Xpix_fin, Xpix_inc
20 236 4                      ! Ypix_ini, Ypix_fin, Ypix_inc
400 0.05                      ! Nksample, Final_Range
\end{verbatim}
\normalsize

\subsection{
Parameter file for \texttt{hfs\_nh3\_cube\_sp} and \texttt{hfs\_nh3\_cube\_mp}}
\small
\begin{verbatim}
"l1287_11_Kv.fits"            ! (1,1) input NH3(1,1) data cube file
0.30  4.0                     ! (1,1) rms, minimum SNR to analyse the spectrum
3                             ! Number of components
27 29  30 32  33 35           ! (1,1) range of channels to search for peak
"l1287_22_Kv.fits"            ! (2,2) input NH3(2,2) data cube file
0.30  4.0                     ! (2,2) rms, minimum SNR to analyse the spectrum
0                             ! Hanning filter half-width (chan): 0=no; >0=yes
1                             ! Boxcar smoothing radius (pixels): 0=no; >0=yes
128 384 3                     ! (1,1)&(2,2) Xpix_ini, Xpix_fin, Xpix_inc
128 384 3                     ! (1,1)&(2,2) Ypix_ini, Ypix_fin, Ypix_inc
400 0.05                      ! Nksample, Final_Range
\end{verbatim}
\normalsize

\section{G. Examples of \hfs\ output files}
\label{aoutput}

\subsection{Header and first lines of \texttt{<source>.synt}}
\label{asynt}

\small
\begin{verbatim}
!Created by HfS_nh3
!DATE       = 2016/01/12
!TIME       = 12:59:00
!TRANSITION = NH3(2,2)
!NCHAN      =          120
!DVCHAN     =      0.30851
!NCOMP      =            2
!DVLINE__1  =      1.05310  
!VLSR____1  =    -18.38200  
!A*TAU_M_1  =      0.00226  
!TAU_M___1  =      0.13445  
!DVLINE__2  =      0.00000  
!VLSR____2  =      0.00000  
!A*TAU_M_2  =      0.00000  
!TAU_M___2  =      0.00000  
!   VELOCITY   SYNTHETIC      COMP_1      COMP_2
   -36.50216     0.00002     0.00002     0.00000
   -36.19365     0.00006     0.00006     0.00000
   -35.88514     0.00011     0.00011     0.00000
   -35.57662     0.00014     0.00014     0.00000
   -35.26811     0.00011     0.00011     0.00000
   -34.95960     0.00006     0.00006     0.00000
   -34.65109     0.00002     0.00002     0.00000
   -34.34258     0.00000     0.00000     0.00000
\end{verbatim}
\normalsize

\subsection{
Header and first lines of \texttt{<source>\_<xoffset>\_<yoffset>.spec}}
\label{aspec}

\small
\begin{verbatim}
!Created by HfS_view
!DATE           = 2016/01/12
!TIME           = 12:59:00
!FITS_FILE      = test1.fits
!NCHAN          =          124
!DVCHAN         =      0.30888
!HANNING_HWDTH  =            0
!BOXCAR_RADIUS  =            0
!X_PIXEL        =           30
!Y_PIXEL        =           40
!   VELOCITY   INTENSITY
   -36.83265    -0.00740
   -36.52377    -0.00101
   -36.21490     0.00281
   -35.90602     0.00570
   -35.59714    -0.00071
   -35.28827     0.00164
   -34.97939    -0.00221
   -34.67051    -0.00472
   -34.36163     0.00010
   -34.05275    -0.00056
   -33.74388    -0.00011
   -33.43500    -0.00030
   -33.12612     0.00054
\end{verbatim}
\normalsize

\subsection{\texttt{<parfile>\_comp\#.out}}
\label{aout}

\scriptsize
\begin{verbatim}
!Created by HfS_cube
!DATE                = 2016/01/12
!TIME                = 13:20:56
!PAR_FILE            = test2_cube.par                                                  
!FITS_FILE           = test1.fits                                                      
!TRANSITION          = NH3(1,1)                        
!TAU_TOT/TAU_M       =     2.0000
!NCOMP               =          2
!COMPONENT           =          1
!VELOCITY_RANGE_MIN  =   -23.3965
!VELOCITY_RANGE_MAX  =   -18.7633
!HANNING_HALF_WIDTH  =          0
!BOXCAR_RADIUS       =         10
!    DELTA_V       ERROR       V_LSR       ERROR     A*TAU_M       ERROR       TAU_M       ERROR         RMS  XOFFSET  YOFFSET  XPIX  YPIX
  8.9086E-01  1.3444E-01 -1.9046E+01  9.7873E-02  1.9160E-02  4.7257E-03  4.8878E+00  1.1600E+00  4.7246E-04   -10.35   -16.56  0079  0041
  7.6377E-01  1.6969E-01 -1.8974E+01  1.2493E-01  2.0565E-02  5.2096E-03  7.2750E+00  1.6474E+00  4.9455E-04   -10.35   -14.49  0079  0044
  7.6430E-01  1.2049E-01 -1.9173E+01  8.5182E-02  1.7090E-02  4.2795E-03  3.7879E+00  9.4660E-01  4.3300E-04   -12.42   -18.63  0082  0038
  8.4162E-01  1.3756E-01 -1.9058E+01  9.8269E-02  1.7949E-02  5.1669E-03  4.6527E+00  1.2671E+00  4.5164E-04   -12.42   -16.56  0082  0041
  8.7368E-01  1.5091E-01 -1.8891E+01  1.0841E-01  1.7925E-02  5.0000E-03  4.8505E+00  1.3107E+00  4.7948E-04   -12.42   -14.49  0082  0044
\end{verbatim}
\normalsize

\end{document}